

\input phyzzx
\def\np{Nucl. Phys.}
\def\pl{Phys. Lett.}

\def\pr{Phys. Rev.}
\def\ap{Ann. Phys.}
\def\cmp{Comm. Math. Phys.}
\def\ijmp{Int. J. Mod. Phys.}

\def\inma{Invent. Math.}
\def\tam{Trans. Am. Math. Soc.}
\def\lmp{Lett. Math. Phys.}
\def\bams{Bull. AMS}
\def\am{Ann. of Math.}
\def\rmp{Rev. Mod. Phys.}
\def\jpsc{J. Phys. Soc. Jap.}
\def\topo{Topology}

\def\half{{1\over 2}}

\def\ex{{\hbox{\rm e}}}

\def\tr{{\hbox{\rm Tr}}}

\def\to{{\rightarrow}}
\def\mani{{\cal M}}

\def\cox{{c_{\hbox{\sevenrm A}}}}

\def\erretres{{{\bf R}^3}}


\def\ifig#1#2#3#4#5{\midinsert\vskip #4truecm \noindent
      \hskip#5truecm \special{postscriptfile #3 scaled
1000}\vskip.3truecm
\narrower\narrower\noindent{\par\begingroup
\noindent\tenpoint\baselineskip1pt {\it Fig.} #1:\
#2\par\endgroup}\endinsert}

\def\ifigc#1#2#3#4#5{\midinsert\vskip #4truecm \noindent
   \hskip#5truecm {\special{postscriptfile #3  scaled
1000}}\vskip.3truecm
        \centerline{{\tenpoint {\it Fig.}  #1:\ #2}} \endinsert}

\tolerance=500000
\overfullrule=0pt

\pubnum={US-FT-8/94\cr hep-th/9407076}
\date={June, 1994}
\pubtype={}
\titlepage

\title{NUMERICAL KNOT INVARIANTS OF FINITE TYPE FROM CHERN-SIMONS
PERTURBATION THEORY}
\author{M. Alvarez and J.M.F. Labastida\foot{e-mail:
LABASTIDA@GAES.USC.ES} }
\address{Departamento de F\'\i sica de Part\'\i culas\break Universidade
de Santiago\break E-15706 Santiago de Compostela, Spain}

\abstract{Chern-Simons gauge theory for compact semisimple groups is
analyzed from a perturbation theory point of view. The general form of
the perturbative series expansion of a Wilson line is presented in terms
of the Casimir operators of the gauge group. From this expansion new
numerical knot invariants are obtained. These knot invariants turn out
to be of finite type (Vassiliev invariants), and to possess an integral
representation.
Using known results about Jones, HOMFLY, Kauffman and Akutsu-Wadati
polynomial
invariants these new knot invariants are computed up to type six for all
prime knots up to six crossings. Our results suggest that these knot
invariants  can be normalized in such a way that they are
integer-valued.}

\endpage
\pagenumber=1

\chapter{Introduction}

Chern-Simons gauge theory \REF\witCS{E. Witten
\journal\cmp&121(89)351} [\witCS] has been studied using non-perturbative
as well as perturbative methods. A variety of non-perturbative studies
have been carried out
\REF\bos{M. Bos and V.P. Nair \journal\pl&B223(89)61
\journal\ijmp&A5(90)959}
\REF\mur{H. Murayama, ``Explicit Quantization of the Chern-Simons
Action", University of Tokyo preprint, UT-542, March 1989}
\REF\lr{J.M.F. Labastida and A.V. Ramallo \journal\pl&B227(89)92
\journal\pl&B228(89)214}
\REF\emss{S. Elitzur, G. Moore, A. Schwimmer and N. Seiberg
\journal\np&B326(89)108}
\REF\fk{J. Frohlich and C. King \journal\cmp&126(89)167}
\REF\djt{G.V. Dunne, R. Jackiw, and C.A. Trugenberger
\journal\ap&194(89)197}
\REF\pietra{S. Axelrod, S. Della Pietra and E. Witten, {\sl J. Diff.
Geom.} {\bf 33} (1991) 787}
\REF\llr{J.M.F. Labastida, P.M. Llatas and A.V. Ramallo
\journal\np&B348(91)651}
\REF\gaku{K. Gawedzki and A. Kupiainen \journal\cmp&135(91)531}
[\bos-\gaku], which has led to many exact results related to polynomial
invariants for knots and links. These include, on the one hand, a general
approach to compute observables related to knots, links and graphs
\REF\witGR{E. Witten \journal\np&B322(89)629}
\REF\martin{S.P. Martin \journal\np&B338(90)244} [\witGR,\martin], and, on
the other hand,  more explicit applications as the computation of
invariants for torus knots and links for the fundamental representation
of the group $SU(N)$
\REF\homfly{J.M.F. Labastida and M. Mari\~no, Santiago preprint,
hep-th/9402093, February 1994} [\homfly] and for  arbitrary
representations of $SU(2)$
\REF\poli{J.M. Isidro, J.M.F. Labastida and A.V. Ramallo
\journal\np&B398(93)187} [\poli], the development of skein rules in a
variety of situations
\REF\ygw{K. Yamagishi, M.-L. Ge and Y.-S. Wu \journal\lmp&19(90)15}
\REF\muk{S. Mukhi, ``Skein Relations and Braiding in Topological Gauge
Theory", Tata preprint, TIFR/TH/98-39, June 1989}
\REF\horne{J.H. Horne \journal\np&B334(90)669} [\ygw,\muk,\horne], and
the computation of invariants in more general cases
\REF\kguno{R.K. Kaul, T.R. Govindarajan\journal\np&B380(92)293}
\REF\kgdos{R.K. Kaul and T.R.  Govindarajan\journal\np&B393(93)392
\journal\np&B402(93)548} [\kguno,\kgdos]. All these studies cover the
analysis of Jones
\REF\jones{V.F.R. Jones \journal\bams&12(85)103}
\REF\jonesAM{V.F.R. Jones\journal\am &  126  (87) 335}
[\jones,\jonesAM],  HOMFLY
\REF\homflyp{P. Freyd, D. Yetter, J. Hoste, W.B.R. Lickorish, K. Millet
and A. Ocneanu \journal\bams&12(85)239} [\homflyp,\jonesAM] and Kauffman
\REF\kauf{L.H. Kauffman\journal\tam&318(90)417} [\kauf]
 polynomials as well as Akutsu-Wadati polynomials
\REF\aw{Y. Akutsu and M. Wadati
\journal\jpsc&56(87)839;\journal\jpsc&56(87)3039  Y. Akutsu, T. Deguchi
and M. Wadati
\journal\jpsc&56(87)3464;\journal\jpsc&57(88)757; for a review see M.
Wadati, T. Deguchi and Y. Akutsu, Phys. Rep. {\bf 180} (1989) 247} [\aw]
for some sets of knots and links.  Perturbative studies of Chern-Simons
gauge theory
\REF\prao{R.D. Prisarski and S. Rao\journal\pr&D22(85)208}
\REF\gmm{E. Guadagnini, M. Martellini and M.
Mintchev\journal\pl&B227(89)111 \journal\pl&B228(89)489}
\REF\gmmdos{E. Guadagnini, M. Martellini and M.
Mintchev\journal\np&B330(90)575}
\REF\alr{L. Alvarez-Gaum\'e, J.M.F. Labastida,  and A.V. Ramallo
\journal\np&B334(90)103, {\sl \np} {B} (Proc. Suppl.) {\bf 18B} (1990)
1}
\REF\brt{D. Birmingham, M. Rakowski and G. Thompson
\journal\np&B234(90)83}
\REF\sor{F. Delduc, F. Gieres and S.P. Sorella\journal\pl&B225(89)367}
\REF\dr{P.H. Damgaard and V.O. Rivelles\journal\pl&B245(89)48}
\REF\cpm{C.P. Mart\'\i n
\journal\pl&B241(90)513}
\REF\aso{M. Asorey and F. Falceto
\journal\pl&B241(90)31}
\REF\bc{A. Blasi and R. Collina
\journal\np&B345(90)472}
\REF\dgs{F. Delduc, C. Lucchesi, O. Piguet and S.P. Sorella
\journal\np&B346(90)513}
\REF\dd{D. Daniel and N. Dorey\journal\pl&B246(90)82}
\REF\dor{N. Dorey\journal\pl&B246(90)87}
\REF\cpmdos{C.P. Mart\'\i n\journal\pl&B263(91)69}
\REF\cpmdosloops{G. Giavarini, C.P. Mart\'\i n and F. Ruiz
Ruiz\journal\np&B381(92)222}
\REF\drwi{D. Bar-Natan and E. Witten\journal\cmp&141(91)423}
\REF\drort{D. Bar-Natan, ``Perturbative aspects of Chern-Simons
topological quantum field theory", Ph.D.  thesis, Princeton Univ., June
1991, Dept. of Mathematics}
\REF\axel{S. Axelrod and I.M. Singer, MIT preprint, October 1991}
[\gmm-\axel] have provided  a rich amount of knowledge on the series
expansion corresponding to  Wilson lines. Many of the works in this
respect deal with the problem of finding which is the renormalization
scheme that leads to the exact results. In this paper we will not
address this issue. We will assume that there exist a scheme in which
the quantum corrections to the two and three-point functions account for
the shift obtained in [\witCS] of the Chern-Simons parameter
$k$. The existence of this scheme has been proved to one loop
[\prao,\alr,\cpm,\cpmdos] and to two loops [\cpmdosloops]. In this paper
we will concentrate on the structure of the perturbative series
expansion. We present a procedure to define numerical knot invariants
from the perturbative expansion. These kinds of studies were first made
by the pioneering work [\gmmdos].

In a previous paper \REF\pert{M. Alvarez and J.M.F.
Labastida\journal\np&B395(93)198} [\pert] we analyzed Chern-Simons gauge
theory from a perturbative point of view. The aim of this paper is to
push forward that analysis to construct new numerical knot invariants.
The main result obtained in [\pert] was the identification of all the
Feynman diagrams of the pertubative series expansion of the vacuum
expectation value of a Wilson line which contribute to its framing
dependence. It was shown that the contribution from all those diagrams
factorizes in the form predicted by Witten [\witCS]. In this paper we
attempt to organize the rest of the perturbative contributions in such a
way that an infinite sequence of numerical knot invariants will be
attached to a given knot.

A Feynman diagram associated to the vacuum expectation value of a Wilson
line in Chern-Simons gauge theory provides a contribution which is the
product of two factors times a power of the coupling constant
$g \sim 1/\sqrt{k}$. The power of this constant characterizes the order
of the Feynman diagram. One of the two factors depends on the gauge
group and the representation chosen for the Wilson line. Its form is
dictated by the Feynman rules. We will denote this factor as group
factor. The second factor corresponds to a series of line and
three-dimensional integrals of a certain integrand also dictated by the
Feynman rules. We will denote this factor as the geometrical factor. The
important point to remark is that given a Feynman diagram the group
factor is independent of the closed curve corresponding to
the Wilson line. On the other hand, the geometrical factor is
independent of the group and representation chosen.

The idea behind the construction of the numerical knot invariants
presented in this paper is the following. Let us consider all Feynman
diagrams associated to the vacuum expectation value of a Wilson line at
a given order in perturbation theory, except those which contribute to
the framing dependence. These diagrams provide a set of group factors.
Among these group factors there is a set of independent ones, \ie, all
the rest can be written as linear combinations of the ones in this set.
The perturbative contribution at the order considered can be written as
a sum of a series of numerical factors times the independent group
factors. Since the whole contribution at a given order is a topological
invariant and the group factors are chosen to be the independent ones,
the numerical factors which enter the contribution are numerical
invariants associated to the knot. These numerical knot invariants can
be regarded as the independent  geometrical factors. The number of
independent group factors at each order in perturbation theory is
finite. Therefore, these procedure allows to assign to each knot a
finite set of numerical invariants at each order. This allows to
associate a numerical sequence to each knot.

An important feature of these numerical knot invariants is that they are
invariants of finite type or Vassiliev invariants
\REF\vassi{V.A. Vassiliev, ``Cohomology of knot spaces"  in {\it Theory
of Singularities and its applications}, (V.I. Arnold, ed.), Amer. Math.
Soc., Providence, RI, 1990, 23}
\REF\vassidos{V.A. Vassiliev, ``Topology of complements to discriminants
and loop spaces"  in {\it Theory of Singularities and its applications},
(V.I. Arnold, ed.), Amer. Math. Soc., Providence, RI, 1990, 23}
\REF\vassitres{V.A. Vassiliev, ``Complements of discriminants of smooth
maps: topology and applications", Translations of Mathematical Monographs,
vol. 98, AMS, 1992}
[\vassi,\vassidos,\vassitres]. This will be shown using recent results
on the
connection between polynomial knot invariants and Vassiliev invariants
by Bar-Natan
\REF\drorcon{D. Bar-Natan, ``Weights of Feynman diagrams and the
Vassiliev knot invariants", preprint, 1991} [\drorcon] and by
Birman and Lin
\REF\birlin{J.S. Birman and X.S. Lin\journal\inma&111(93)225}
\REF\lin{X.S. Lin, Vertex models, quantum groups and Vassiliev knot
invariants, Columbia University preprint, 1991} [\birlin,\lin]. An
important object associated to a Vassiliev invariant is its actuality
table [\vassi,\vassidos,\birlin,\lin].  At a given order there are
several invariants which have as type their order in perturbation
theory. These invariants generate sets of actuality tables. There can
not be more independent tables than the dimension of the space of
Vassiliev invariants of the given type. To the order studied this is
consistent with our results. Another important feature of the numerical
knot invariants we are dealing with is that there seems to exist a
normalization such that these knot invariants are integer-valued.

The new numerical knot invariants presented in this paper are framing
independent, possess integral expressions, and are integers when properly
normalized. Their integral expressions are rather cumbersome and
therefore, in general, they are hard to compute. There is, however, an
alternative way to obtain these invariants using exact results for knot
polynomial invariants. Since the numerical invariants are universal in
the sense that they are independent of the group and representation
chosen, one may obtain sets of linear equations for them comparing the
perturbative series expansion dictated by Chern-Simons gauge theory to
known exact results. We apply this method in this work and we present
the computation of the new numerical invariants for all prime knots up
to six crossings to order six using known polynomial invariants. The
results are summarized in Table I.

The paper is organized as follows. In sect. 2 we review the results of
[\pert] and we summarize the Feynman rules of the theory. In sect. 3 the
general  form of the perturbative series expansion is given and the
complete details are worked out up to order six. In sect. 4 the
numerical knot invariants are defined and their features  are analyzed;
in particular, it is proven that they are of finite type. In sect. 5
these invariants are computed up to order six for all prime knots up to
six crossings. Finally, in sect. 6 we state our final remarks. There are
in addition three appendices. Appendix A contains our group theoretical
conventions and the description of the calculation of Casimirs. Appendix
B deals with the discussion of some technical details regarding the
analysis of the general structure of the perturbative series expansion.
In appendix C we present a summary of known polynomial invariants which
are used in the calculations carried out in sect. 5.

\endpage

\chapter{Perturbative Chern-Simons gauge theory, Feynman rules and
factorization theorem}

In this section we will present first a brief review of Chern-Simons
gauge theory from a perturbation theory point of view for a general
compact semisimple group $G$. Our approach uses standard  perturbative
quantum field theory, and utilizes Feynman diagrams as the main tool. We
apologize if this brief review  occasionally becomes too explicit for a
field theorist but we expect that the details will become useful for
people more mathematically oriented.

Let us consider a $G$ gauge connection
$A$ on a compact boundaryless three-dimensional manifold $\mani$.  The
Chern-Simons action is defined as,
$$ S_0(A) = {k\over 4\pi}\int_{\mani} \tr(A\wedge d A + {2\over 3}
A\wedge A\wedge A),
\eqn\cinco
$$ where $k$ is an arbitrary positive integer. The symbol ``Tr" denotes
the trace in the  fundamental representation of $G$. Notice that the
action \cinco\ does not depend on the metric on
$\mani$. In defining the theory from a perturbation theory point of view
we must give a meaning to vacuum expectation values of operators, \ie, to
quantities of the form,
$$
\langle {\cal O} \rangle = {1\over Z}
\int [D A] {\cal O}(A) \exp\Big(iS(A)\Big),
\eqn\seis
$$ where $Z$ is the partition function,
$$ Z=\int [D A]  \exp\Big(iS(A)\Big).
\eqn\seisp
$$ In \seis\ ${\cal O}(A)$ is a function of $A$ which might be local or
non-local. The integrals entering  \seis\ and \seisp\ are functional
integrals. We do not aim to a rigorous definition of these objects in
terms of a measure, but to a exposition of the perturbative analysis of
Chern-Simons theory. In this context these formal definitions are
accurate enough.  In order to obtain topological invariants, the
operators entering
\seis\ are chosen to be gauge invariant operators which do not depend on
the three-dimensional metric. These operators are related to knots,
links and graphs \REF\witGR{E. Witten \journal\np&B322(89)629}
[\witCS,\witGR]. The requirement of gauge invariance comes from the fact
that the exponential in \seis\ is invariant under gauge transformations
of the form,
$$ A_\mu \to h^{-1}A_\mu h + h^{-1}\partial_\mu h,
\eqn\siete
$$ where $h$ is an arbitrary continuous map $h:\mani\to G$. This implies
that the integration over gauge connections has to be restricted to an
integration over gauge connections modulo gauge transformations. Our
choice of gauge fixing will be the same as the one taken  in
[\alr,\pert].

Let us redefine the constant $k$ and the field $A$ in such a way that
the action \cinco\ becomes standard from a perturbation theory point of
view. Defining
$$ g = \sqrt{{4\pi\over k}},
\eqn\ocho
$$ one finds, after rescaling the gauge connection,
$$ A_\mu \to g A_\mu,
\eqn\nueve
$$ that the  Chern-Simons action takes the form:
$$ S(A) = \int_{\mani} \tr(A\wedge d A + {2\over 3} g A\wedge A\wedge A).
\eqn\diez
$$ If we choose a trivialization for the tangent bundle of the three
manifold
$\mani$, the previous action can be written in components. Although the
tangent bundle to
$\mani$ can be trivialized, this can be done in infinitely many ways.
This is the origin of a phase ambiguity in $Z$ which is discussed, for
example, in [\witCS], but this problem is immaterial for us. Following
the group-theoretical conventions stated in Appendix A, the action in
components reads,
$$ S(A)= {1\over 2}\int_{\mani}
\epsilon^{\mu\nu\rho}\Big(A_{\mu}^a\partial_{\nu}A_{\rho}^a-{g\over
3}f^{abc}A_{\mu}^a A_{\nu}^b A_{\rho}^c\Big).
\eqn\actioncomp
$$

The standard procedure to compute \seis\ and \seisp\ from a perturbation
theory point of  view involves the introduction of a source function $J$
for the gauge field $A$ and a modification of the action in the form,
$$ S(A,J)= \int_{\mani}\Big[
\half\epsilon^{\mu\nu\rho}\Big(A_{\mu}^a\partial_{\nu}A_{\rho}^a-{g\over
3}f^{abc}A_{\mu}^a A_{\nu}^b A_{\rho}^c\Big)+J^{\mu\,\,a}A_{\mu}^a \Big].
\eqn\actionsource
$$ The original vacuum expectation value \seis\ and the partition
function \seisp\ are recovered setting $J=0$. Standard arguments in
quantum field theory allow to write,
$$
\langle {\cal O} \rangle = \langle {\cal O}(J) \rangle \big|_{J=0},
\eqn\pera
$$ where,
$$
\eqalign{
\langle {\cal O}(J) \rangle & =  {1\over Z[J]} \int [D A]  \exp(iS(A,J))
\cr &= { {\cal O}({\delta\over \delta J})
\exp\Big{\{}{-g\over
6}\int_{\mani}\epsilon^{\mu\nu\rho}f^{abc}{\delta\over\delta
J_{\mu}^a}{\delta\over\delta J_{\nu}^b} {\delta\over\delta
J_{\rho}^c}\Big{\}}
\exp\Big{\{}{i\over 2}\int_{\mani}\int_{\mani} J_{\sigma}^d
D^{\sigma\tau}_{de}J_{\tau}^e\Big{\}}
\over
\exp\Big{\{}{-g\over
6}\int_{\mani}\epsilon^{\mu\nu\rho}f^{abc}{\delta\over\delta
J_{\mu}^a}{\delta\over\delta J_{\nu}^b} {\delta\over\delta
J_{\rho}^c}\Big{\}}
\exp\Big{\{}{i\over 2}\int_{\mani}\int_{\mani} J_{\sigma}^d
D^{\sigma\tau}_{de}J_{\tau}^e\Big{\}}  }. \cr}
\eqn\meloco
$$ In this last expression $D^{\mu\nu}_{ab}$ represents the propagator or
two-point function at tree level. Its explicit form is  given below. The
expansion of this expression in powers of $g$ leads to the standard
perturbative series expansion in quantum field theory. The best way to
organize the different contributions is to use Feynman diagrams. These
are obtained from the Feynman rules which can be read in part from
\meloco. To complete the set of Feynman rules one must be more specific
about the operators  ${\cal O}(A)$.  The presence of the denominator in
\meloco\ has a simple interpretation in terms of Feynman diagrams: one
must take only those diagrams which are connected.

Before entering into the discussion on the structure of \meloco\ we must
first introduce the operators which will be of interest for us.  We will
refer to these as observables. As in any gauge theory, the observables
have to be gauge invariant. In the case of a topological theory, as the
one at hand, we also would like to have observables without any
dependence on the metric of the manifold $\mani$. These conditions are
satisfied for the so-called Wilson loops. To introduce these
objects, let us recall the notion of non-Abelian holonomy. If $C$ is a
parametrized loop in $\mani$, then given a $G$ connection $A$ on
$\mani$, carrying a representation $R$ of $G$, and any two points $C(s)$
and $C(t)$ on the loop, we  define an element $W^R_C(s,t)$  of $G$ in
the following fashion,
$$
W^R_C(s,t)={\rm{P}}\exp\Big{\{}g\int_{C(s)}^{C(t)}A\Big{\}},
\eqn\uvedoble
$$
where P stands for ``path ordered''. This concept is analogous to the
concept of time ordering in quantum field theory, and can be briefly
described as follows: in the expansion of the exponential some products
of connections $A(s)$ defined at different points $C(s)$ of the path
will appear. The path ordering puts these factors in decreasing order of
the parameter $s$. It is this ordering what makes the product of two
such $A(s)$ to be equivalent to the above defined two-point function.
For a thorough exposition of these and other common concepts in quantum
field theory, see, for example,
\REF\itzyk{C. Itzykson and J.B. Zuber, {\it Quantum Field Theory},
McGraw-Hill, 1980} [\itzyk].

We shall denote the holonomy, or the parallel transport around the loop,
by $W^R_C(s)\equiv W^R_C(s,s)$. This $W^R_C(s)$ is an element of $G$,
and the Wilson loop is defined to be simply
the trace of the holonomy of the connection 1-form
$A$ along $C$,
$$ W^R_C=\tr\Big({\rm{P}}\exp\Big{\{}g\oint_C A\Big{\}}\Big).
\eqn\holonomy
$$ Notice that we have dropped the dependence on the initial point $s$.
The trace is taken over the representation
$R$ of the algebra of $G$ carried by the connection $A$. The two chief
features of
\holonomy\ are its gauge invariance and its independence on any metric
whatsoever. These attributes single out the Wilson lines as the best
candidates for observables in a gauge invariant topological field
theory. Some generalizations of these objects are defined in [\witGR],
but will not be considered in this work. The connection of the Wilson
lines with knot theory, discovered by Witten in [\witCS], is established
through the dependence of
$W^R_C$ on the loop $C$. This loop can be knotted in any fashion, and
the vacuum expectation value of
$W^R_C$ is related to knot invariants.

The vacuum expectation values of the Wilson line are defined using
\seis\ and taking
\holonomy\ as the operator $\langle {\cal O} \rangle$. We will restrict
ourselves in  the rest of this work to the three-manifold $\erretres$
(so that effectively one is dealing with knot invariants on $S^3$). In
the perturbative expansion one finds, besides convolutions as dictated
by the first two Feynman rules of
\FIG\figfeynman\  Fig. \figfeynman, traces of generators of the algebra
$G$ in the representation $R$. This fact introduces the need for an
extra Feynman rule reflecting the attachment of the gauge field $A$ to
the Wilson line. This rule is the third one depicted in Fig. \figfeynman.
These traces, together with part of the other two Feynman rules,
generate group factors. For a given order in the perturbative expansion,
the group-theoretical factors and the convolutions factorize and can be
calculated independently. This fact will be of some importance in what
follows, since our organization of the perturbative series is guided by
the structure of the group-theoretical factors. The first Feynman rule
in Fig.
\figfeynman\ involves the propagator $D_{ab}^{\mu\nu}(x-y)$, while the
second involves the vertex
$V_{abc}^{\mu\nu\rho}(x,y,z)$ or three-point function at tree level.

Actually the perturbative expansion is divergent in the sense that the
convolutions of propagators and vertices are divergent integrals which
need to be regularized. Moreover, the gauge invariance of \cinco\
indicates that the functional integrals have to be restricted to
integrals over gauge connections modulo gauge transformations. This last
issue can be solved by introducing some unphysical fields (ghosts)
in the action. These two problems have been thoroughly examined in the
last years [\gmm-\axel]. In this paper we will follow the approach taken
in [\alr], where a Pauli-Villars regularization was introduced. We do
not give the Feynman rules corresponding to ghost and Pauli-Villars
fields since these fields only enter in loops and we will take the
results obtained in [\alr] for one-loop Green functions. These results
are summarized in  \FIG\figresults\ Fig. \figresults. As stated in the
introduction we will further assume that there exist a scheme in which
the quantum corrections to the two and three-point functions account for
the shift obtained in [\witCS] of the Chern-Simons parameter
$k$. The existence of this scheme has been proved to one loop
[\prao,\alr,\cpm,\cpmdos] and to two loops [\cpmdosloops]. This implies
that we do not have to worry about Feynman diagrams containing two and
three-point functions at one or higher loops.

Another important contribution inherited in the vacuum expectation value
of a Wilson line is the framing factor. In our previous work [\pert] all
diagrams contributing to this factor were identified. We will make a
brief review of that result in the rest of this section. To carry this
out we must introduce the following classification of propagators,
which, on the other hand,  will be also useful in other sections of the
paper. We call ``free'' those propagators with both endpoints on the
Wilson line, and ``collapsible'' those free propagators whose endpoints
can get together without crossing over any point belonging to other
subdiagram. For example, diagram c of \FIG\figvarios\ Fig. \figvarios\
contains two free and one collapsible propagators.

The main result of [\pert] is the factorization theorem, which enables
us to identify the Feynman diagrams that contribute to the framing
dependence of the Wilson line. This is essential to our approach in two
senses. First, the numerical knot invariants we are going to present are
based on the idea that the knot invariants should not depend on the
framing, which is not intrinsic to the knot, and therefore all framing
dependent contributions should be isolated and discarded. Second, it
explains why in other approaches to Vassiliev invariants similar types
of diagrams have to be  set to zero [\drort].

To state the factorization theorem we need to introduce some notation.
We will be considering  diagrams corresponding to a given order
$g^{2m}$ in the perturbative expansion of a knot, and to a given number
of points running over it, namely $n$. We will denote by
$\{i_1,i_2,\ldots,i_n\}$ a  domain of integration where the order of
integration is $i_1<i_2<...<i_n$, being
$i_1,i_2,\dots,i_n$ the points on the knot (notice the condensed
notation) where the internal lines of the  diagram are attached. The
integrand corresponding to that diagram will be denoted as
$f(i_1,i_2,\ldots,i_n)$. Diagrams are  in general composed of
subdiagrams, which may be connected or non-connected. For a given
diagram we can make specific choices of subdiagrams depending on the
type of factorization which is intended to achieve. For example, for a
diagram like $c$ of  Fig. \figvarios\ one may choose as subdiagrams the
three free propagators, or one may choose a subdiagram to be the
collapsible  propagator and other subdiagram to be the one built by two
crossed free propagators.

We will consider a set of diagrams ${\cal N}$ corresponding to a given
order $g^{2m}$, to a given number of points attached to the knot,
$n$, and to a given kind. By kind we mean all diagrams containing
$n_i$  subdiagrams of type $i$, $i=1,...,T$. By $p_i$ we will denote the
number of points which a subdiagram of type $i$ has attached to the
knot. For example, if one considers diagrams at order $g^6$ with $n=6$
points attached to the knot, with three subdiagrams which are just  free
propagators, this set is made out of diagrams $a$ to $e$ of Fig.
\figvarios.\ However, if one considers diagrams at order $g^6$ with
$n=6$ with a subdiagram consisting of a free propagator and a triple
vertex, this set is made out of diagrams $f$ and $g$.  The contribution
from all diagrams in
${\cal N}$ can be written as the following sum:
$$
\sum_{\sigma\in \Pi_n} \oint_{i_1,i_2,\ldots,i_n}
f(i_{\sigma(1)},i_{\sigma^(2)},\ldots,i_{\sigma(n)}),
\eqn\newuno
$$ where $\sigma\in \Pi_n$, being $\Pi_n\subset P_n$ a subset of the
symmetric group of $n$ elements. Notice that $\Pi_n$ reflects the
different shapes of the diagrams in ${\cal N}$. In \newuno\ the
integration region has been left fixed for all the diagrams and one has
introduced  different integrands. One could have taken the opposite
choice, namely, one could have left fixed the integrand and sum over the
different domains associated to ${\cal N}$. The first statement
regarding the factorization theorem just refers to these two possible
choices. Let us define the domain resulting of permuting
$\{i_1,i_2,\ldots,i_n\}$ by an element $\sigma$ of the symmetric group
$P_n$ by
$$ {\cal
D}_\sigma=\{\,i_{\sigma(1)},i_{\sigma(2)},\ldots,i_{\sigma(n)}\,\},
\eqn\funo
$$ then the following result immediately follows.

Statement 1: {\sl The contribution to the Wilson line of the sum of
diagrams whose integrands are of the form:
$$ f(i_{\sigma (1)}, i_{\sigma(2)}, \dots, i_{\sigma(n)}),
\eqn\ftres
$$  where $\sigma$ runs over a given subset $\Pi_n \in P_n$ with a
common domain of integration is equal to the sum of the integral of
$f(i_1,i_2,\ldots,i_n)$ over
${\cal D}_\sigma$ where $\sigma \in \Pi_n^{-1}$:
$$
\oint_{i_1,i_2,\ldots,i_n} \sum_{\sigma \in \Pi_n}
f(i_{\sigma(1)},i_{\sigma^(2)},\ldots,i_{\sigma(n)}) =
\sum_{\sigma \in \Pi_n^{-1}} \,\,\,\,\, \oint_{{\cal D}_\sigma}
f(i_1,i_2,\ldots,i_n).
\eqn\fcuatro
$$}

The idea behind the factorization theorem is to organize the diagrams in
${\cal N}$ in such a way that one is summing over all possible
permutations of domains. Summing over all domains implies that one can
consider the integration over the points corresponding to each
subdiagram as independent and therefore one can factorize the
contribution into a product given by the integrations of each subdiagram
independently. This leads to the following statement.

Statement 2: (Factorization theorem) {\sl Let $\Pi_n '$ be the set of
all possible permutations of the domains of integration of diagrams
containing subdiagrams of types $i=1,...,T.$
If $\, \Pi_n ^{-1} = \Pi_n '$,
the sum of integrals over ${\cal D}_\sigma$, $\sigma \in \Pi_n^{-1}$, is
the product of the integrals of the subdiagrams over the knot, being the
domains all independent,
$$
\sum_{\sigma \in\Pi_n^{-1}}  \,\,\,\,\,
\oint_{{\cal D}_\sigma} f(i_1,\ldots,i_n) =
\prod_{i=1}^{T} \biggl(\oint_{i_1,\ldots,i_{p_i}}
 f_i(i_1,\ldots,i_{p_i}) \biggr)^{n_i}.
\eqn\fcinco
$$ In \fcinco\ $n_i$ denotes the number of subdiagram of type $i$ and
$p_i$ its number of points attached to the knot. }

The proof of this statement is  trivial since having all possible
domains it is clear that one can write the integration considering
subdiagram by subdiagram, the result being the product of all the
partial integrations over subdiagrams.

As a consequence of the factorization theorem we can state two
corollaries about the framing independence of diagrams which do not
contain one-particle irreducible subdiagrams  corresponding to two-point
functions whose endpoints could get together.  These corollaries refer
to any kind of knot. Their statements are:

Framing dependent diagrams:  {\sl A diagram gives a framing dependent
contribution to the perturbative expansion of the knot if and only if it
contains at least one collapsible propagator. Moreover, the order of $m$
in its contribution, the self-linking number,
equals the number of collapsible propagators}.

Factorization of the framing dependence: {\sl  If all the contribution
to the self-energy comes from one loop diagrams, then $\bigl< W_C^R
\bigr> =F(C,R)
\ex^{2\pi i m h_R}$ where $F(C,R)$ is framing independent but knot
dependent, and the exponential is manifestly framing dependent but knot
independent}.
\vskip .5cm

The quantities $h_R$ and $m$ appearing in the framing dependence factor
are, respectively, the conformal weight associated to the representation
$R$, and the integer  which labels the framing (self-linking number). The
standard framing corresponds to $m=0$.  The factorization of the framing
dependence was proven in   [\pert] for $SU(N)$ in the fundamental
representation but it is obvious from the proof that it generalizes for
any representation of any semisimple group. We end this section
recalling that a full account of these results  can be found in [\pert].
They are the cornerstones of our approach to the finite-type invariants
associated to perturbative Chern-Simons theory.

\endpage

\chapter{General structure of the perturbative expansion} In this
section we will analyze the structure of the perturbative series
expansion associated to  Chern-Simons gauge theory with an arbitrary
compact semisimple gauge group $G$. We will discuss the general form of
this series and we will present its exact  form up to order six.  Let us
consider the vacuum expectation value of the Wilson line \holonomy\
corresponding to an arbitrary knot in an arbitrary representation $R$ of
$G$.  The contour integral in \holonomy\ corresponds to any path
diffeomorphic to the knot. To compute the vacuum expectation values of
this operator in perturbation theory we have to consider all diagrams
which are not vacuum diagrams since we consider normalized vacuum
expectation values, \ie, the functional integration where the operator
is inserted is divided by the partition function $Z$ as in
\seis. Also, we will not consider diagrams which include collapsible
propagators because they only contribute to the dependence of the vacuum
expectation value of $W$ on the framing. Finally we can omit the
insertion of loops in every two and three-point subdiagram since, as
stated before, their only effect is to provide the shift $k\to k-\cox$.
We stress these two last points because they greatly simplify the
perturbative series. The framing and the shift are viewed as objects not
intrinsic to the knot and are therefore ignored.

{}From the Feynman rules presented in the previous section follows that
the perturbative expansion of the vacuum expectation value of the Wilson
line operator
\holonomy\ has the form,
$$
\langle W^R_C \rangle = d(R) \sum_{i=0}^\infty
\sum_{j=1}^{d_i}\alpha_{ij}r_{ij} x^i,
\eqn\expansion
$$ where $x={2\pi i\over k}=ig^2/2$ is the expansion parameter,
$d(R)$ is the dimension of the representation $R$,  and
$\alpha_{0,1}=r_{0,1}=1$, $d_0=1$ and $d_1=0$. Notice that we are
dispensing with the shift and the framing factor: only $k$ appears in
the denominator of
$x$ and there is no linear term in the expansion ($d_1=0$)
\foot{At order $x$ there is only the contribution from a diagram
containing one collapsible propagator which according to the results
presented in the previous section corresponds to framing.}.
The factors
$\alpha_{ij}$ and
$r_{ij}$ appearing at each order $i$ incorporate all the dependence
dictated from the Feynman rules apart from the dependence on  the
coupling constant, which is contained in $x$.
Of these two factors, in the $r_{ij}$ all
the group-theoretical dependence is collected. These will be called
group factors. The rest is contained in  the $\alpha_{ij}$. These last
quantities, which will be called geometrical factors, have the form of
integrals over the Wilson line of products of propagators, as dictated
by the Feynman rules. The first index in $\alpha_{ij}$ denotes the order
in the expansion and the second index labels the different geometrical
factors which can contribute at the given order. Similarly, $r_{ij}$
stands for the independent group structures which appear at order
$i$ which are also dictated by the Feynman rules. The object
$d_i$ in \expansion\ will be called the ``dimension'' of the space of
invariants at a given order. In our approach denotes the number of
independent group structures which appear at that order. The main
content of this section is the characterization of these group
structures. Notice that while the geometrical factors $\alpha_{ij}$ are
knot dependent but group and representation independent (and therefore
one must keep in mind their full form $\alpha_{ij}(C)$), the group
factors are group and representation dependent but knot independent
($r_{ij}(R)$).

The group factors which appear in the expansion \expansion\ are group
invariants made out of traces of the generators contracted with the
tensors $\delta_{ab}$ and
$f_{abc}$. These tensors are described in Appendix A. The Lie algebra
(A1) satisfied by the generators and the Jacobi identity (A2) relate
some of the group invariants which appear at a given order. In the
expansion \expansion\ the group factors  $r_{ij}$,
$j=1,\dots,d_i$, are the independent ones at a given order $i$. In
general, these are obtained as follows. First one writes the group
factors of all the diagrams with no collapsible propagator and no two
and three-point one or higher-loop subdiagrams contributing to a given
order $i$. Then one makes use of (A1) and (A2) so that a selected set of
independent group factors is chosen.

The characterization of the independent group factors is carried out in
two steps. First the independent Casimirs are constructed. Then, the
independent group factors are built using these independent Casimir
invariants. Casimirs invariants will be denoted by
$C_i^j$, where the subindex denotes the order of the Casimir and the
superindex labels the different Casimirs that appear at a given order.
It is worth to recall here that the order $i$ of a Casimir is such that
$2i$ equals the number of generators plus the number of structure
constants which appear in its expression. The independent Casimirs for a
semisimple group up to order 6 turn out to be the following:
$$
\eqalign{ C_2' d(R) &=f_{apq}f_{bqp}\tr (T_a T_b), \cr C_3' d(R)
&=f_{apq}f_{bqr}f_{crp}\tr (T_a T_b T_c), \cr C_4 d(R)
&=f_{apq}f_{bqr}f_{crs}f_{dsp}\tr (T_a T_b T_c T_d), \cr C_5 d(R)
&=f_{apq}f_{bqr}f_{crs}f_{dst}f_{etp}\tr (T_a T_b T_c T_d T_e), \cr
C_6^1 d(R) &=f_{apq}f_{bqr}f_{crs}f_{dst}f_{etu}f_{rup}\tr (T_a T_b T_c
T_d T_e T_r), \cr C_6^2 d(R)
&=f_{apq}f_{brs}f_{ctp}f_{dur}f_{eqs}f_{gtu}\tr (T_a T_b T_c T_d T_e
T_g).
\cr}
\eqn\casimirs
$$ Several comments are in order. First, in general, if there is only
one Casimir at a given order we will label it by $C_i$ instead of
$C_i^1$. Second, notice that the first two have been denoted by
$C_2'$ and $C_3'$ instead of  $C_2$ and $C_3$. The reason for this is
that the notation
$C_2$ and $C_3$ will be reserved to these two Casimirs times appropriate
factors. This will simplify the expressions for the group factors.
Third, the factor $d(R)$, the dimension of the representation, is
introduced for convenience. Notice that for  an irreducible
representation there is a trace of the identity matrix on the right hand
side. Fourth, it is at order 6 when two independent Casimirs appear for
the first time. Notice that for a specific semisimple group these two
Casimirs might not be independent. However, in general they are. What is
meant by independence is that
$C_6^1$ can not be written in terms of $C_6^2$ plus terms which are
products of lower order Casimirs making use of (A1) and (A2). The number
of independent Casimirs at  order $i$ will be denoted by $c_i$. We have
$c_2=c_3=c_4=c_5=1$ and $c_6=2$. The specific values of the independent
Casimirs in \casimirs\ are given in Appendix A for the groups $SU(N)$
and $SO(N)$ in their fundamental representations and for $SU(2)$ in an
arbitrary irreducible representation.

As announced above, the second and third order Casimirs will be
redefined for later convenience. First notice that using (A2) and (A5)
one finds,
$$
\eqalign{ C_2' d(R) &= C_A \tr (T_a T_a), \cr C_3' d(R) &= -\half C_A
f_{abc}\tr (T_a T_b T_c), \cr}
\eqn\casimirsdos
$$ where $C_A$ is the quadratic Casimir in the adjoint representation.
On the other hand, as it will become clear below, the two traces on the
right hand side of
\casimirsdos\ are the quantities which more often appear in group
factors. We then define,
$$ C_2 = {1\over d(R)} \tr (T_a T_a)  = C_2' / C_A,
\,\,\,\,\,\,\,\,\,\, C_3 = -{1\over d(R)} f_{abc}\tr (T_a T_b T_c) =
2C_3' / C_A.
\eqn\casimirtres
$$ The diagrams associated to the independent Casimirs
$C_2$, $C_3$, $C_4$, $C_5$, $C_6^1$ and $C_6^2$ are shown in
\FIG\figtresnum\ Fig.\figtresnum. From those diagrams one easily writes
down the Casimirs using only the part of the Feynman rules concerning
group-theoretical factors.

To obtain the group factors $r_{ij}$, $j=1,\dots,d_i$, to a given order
one must consider all the diagrams with no collapsible propagators and
no two and three-point loop-insertions, and write their group factors in
terms of the independent Casimirs and products or ratios of them. We
will present an algorithm  which leads to the independent group
factors at a given order. We will discuss first the case of  simple
groups and then we will generalize  it for the  semisimple case.

We will introduce first some notation. Let us consider an arbitrary
diagram
$D$ of the perturbative expansion. We will denote by  $V$ the number of
vertices, by $P_{\rm{free}}$ the number of free propagators and by $P_I$
the number of ``internal'' propagators with both endpoints attached to
vertices present in a given diagram. It is convenient to define he
number of ``effective'' propagators, $P$, as, $$ P=2V-P_I+P_{\rm{free}}.
\eqn\eff
$$ It is also useful to introduce a grading $p(i)$ for the Casimirs of
order $i$, $C_i^k$, $k=1,\dots,c_i$, equal to the effective number of
propagators corresponding to their associated diagram (see Fig.
\figtresnum).
$$ p(2)=1, \qquad p(3)=2, \qquad p(i)=i \quad {\rm{if}} \quad i\geq 4.
\eqn\grading
$$

The main result leading to the characterization of group factors for the
case of simple groups is the following. Given a diagram $D$ with $P\geq
3$ effective propagators,   its group-theoretical dependence takes the
form,
$$ r(D)=\sum_{\{ S_P
\}}a_{S_P}\prod_{i=2}^P\prod_{k=1}^{c_i}\big(C_i^k\big)^{S(i,k)},
\eqn\ansatz
$$ where the sum runs over all possible sets $\{ S_P \}$ which we are
about to define, and $a_{S_P}$ are some rational numbers depending on
the diagram $D$.  The sets $\{ S_P \}$ are all the possible collections
of integers  $\{S(i,k): 2\leq i\leq P,\quad 1\leq k \leq c_i\}$
satisfying the following conditions:
$$
\eqalign{
\sum_{i=2}^P \sum_{k=1}^{c_i} p(i)S(i,k)=&P, \cr S(i,k)\geq &0 \quad
{\rm{if}}\quad i\neq2, \cr S(2,1)+S(3,1) \geq &0, \cr S(2,1) \geq & 2-P.
\cr}
\eqn\conditions
$$ According to \ansatz, the integers $S(i,k)$ correspond to the number
of times that the subdiagram associated to the
 Casimir $C_i^k$ appears. The meaning of conditions \conditions\
is the  following. The first one simply imposes that the total number of
effective propagators is correct, and the second one requires that for
$i\neq 2$ the subdiagrams corresponding to the  Casimirs $C_i^k$ appear
a positive number of times. The possible negative values of the $S(i,k)$
for $i=2$ are
due to the presence of $C_A$ factors, as explained below. Finally, the
last two inequalities are the constraints on these values. Notice that
we have used the fact that $c_2=c_3=1$. The formula \ansatz\ is valid
for $P\geq 3$. The group structures corresponding to $P=2$ must be
obtained independently. However, as discussed below, these are very
simple. Finally, we must mention that
\ansatz\ provides all the group structures, including the ones
contributing to framing. These will be identified and discarded
thereafter.

The proof of \ansatz\  goes by induction in $P$. Assume that we are
given a diagram $D_p$ with
$P=p$ as the one depicted in \FIG\generaldiag\ Fig. \generaldiag\ and
that its group factor is $r(D_p)$ according to \ansatz.

\noindent Now one more free propagator is introduced, and the key idea is
to move one of its endpoints next to the other by means of the
commutation relations. This produces a series of diagrams with $p+1$
effective propagators. The last element of this series would include a
diagram similar to Fig. \generaldiag\ with a collapsible propagator
included, as the one depicted in
\FIG\freeprop\ Fig. \freeprop. Its group factor will be $C_2 \, r(D_p)$.

According to our general framework, this is a framing dependent
contribution which should be set to zero, but we are interested only in
group factors at this moment. The remaining diagrams can be represented
as the one in \FIG\generalvert\ Fig.\generalvert.

 The next step is based on the observation of the fact that all diagrams
like Fig. \generalvert\ have a vertex with two points attached to the
Wilson line. Again the procedure employs the commutation relations in
order to make these points closer until we finish in a diagram with a
``fishtail'' like the one on the left of
\FIG\generalfish\ Fig. \generalfish. By fishtail we mean a configuration
of internal lines in a diagram such the two  propagators attached
contiguously  to the Wilson line are joined to the same vertex.
The other diagrams generated are similar to  the one on the right in the
same figure.

As explained in Appendix B fishtails amount to a factor $C_A/2=-C_3/C_2$
and therefore the group factor corresponding to the first diagram in
Fig. \generalfish\ can be written as $-(C_3/C_2)\, r(D_p)$. The general
procedure reproduces this scheme. Take one of the vertices generated in
the previous steps such that it has two points connected to the Wilson line
and move them until one gets a fishtail. Then repeat the same routine
with the diagrams arisen due to those movings. The procedure finishes
when one is left with diagrams such that all vertices have at most one
point on the Wilson line. This describes a diagram with closed loops
which can be connected or disconnected, as is schematically shown in
\FIG\casimirlike\ Fig.
\casimirlike. In the first case it can be written as a combination of
the Casimirs in
$C_{p+1}^i$ and in the second as a combination of products of lower
Casimirs with a total number of effective propagators equal to $p+1$.
These diagrams will be referred to as ``Casimir-like''.

The result is that the group factors needed for $r(D_{p+1})$ are
constructed by multiplying those of $r(D_p)$ by $C_2$ or $C_3/C_2$, by
considering products of lower Casimirs with a total number of effective
propagators equal to $p+1$, and by including the new Casimirs
$C_{p+1}^k$. In our notation this means that $P$ can be increased in one
unit only by some of the following procedures: adding $1$ to $S(2,1)$;
adding $1$ to $S(3,1)$ and
$-1$ to
$S(2,1)$; giving values to the $S(i,k)$ corresponding to lower Casimirs
in such a way that the number of effective propagators equals $p+1$;
finally, setting
$S(p+1,k)=1$ (only one $k$ at a time) for the new Casimirs. This is
verified by
\ansatz\ for $P=p+1$ if the numbers $S(i,k)$ satisfy \conditions.

To end the proof of \ansatz\ we must present the group structures for
the case $P=3$.  Let us  discuss first, as promised, the ones for $P=2$.
This offers no difficulty because this case is nearly trivial. The
diagrams contributing  for $P=2$ are the ones in
\FIG\casson\ Fig. \casson\  plus the diagram containing two collapsible
propagators, which has not been pictured. The calculation of the  group
factors corresponding to these diagrams is rather straightforward and
they turn out to be $(C_2)^2$ and $C_3$. For $P=3$ one has group factors
from the diagrams presented in Fig. \figvarios. Although a larger set
than in the $P=2$ case, their group factors are also computed very
easily.  The independent ones are:
$$
\eqalign{ P=3:& \qquad (3,0)\quad \longrightarrow \big(\sum C_2\big)^3
\cr
               & \qquad (1,1)\quad \longrightarrow \big(\sum C_2\big)
\sum C_3 \cr
               & \qquad (-1,2)\quad \longrightarrow \sum
\big[\big(C_2\big)^{-1}\big(C_3\big)^2\big]
\quad\ast\cr}
$$ where we have used the symbol
$\sum C_i^k$ as a shorthand for $\sum_{l=1}^n C_i^{k\,(l)}$. A similar
condensed notation will be adopted in the examples presented below.  An
asterisk signals the group factors which contribute to the framing
independent part. It follows from the structure of the proof that these
will be all group factors which do not contain positive powers of $\sum
C_2$. The numerical sequence written on the left of the group factors
correspond to the set of integers $S(i,k)$. Clearly, they satisfy
\conditions\ and  the proof of \ansatz\ is completed.

The generalization to the semisimple case can be treated as follows. It
may happen that different diagrams lead to the same group structure in
the simple case. This is explained in  Appendix B.  In \ansatz\ the
products of Casimirs can sometimes be separated in subproducts in such a
way that each subproduct corresponds to a possible subdiagram. These
decompositions would yield all diagrams which in the simple case have
the same group factor. If the algebra is now
$A=\oplus_{l=1}^n A_l$ the Casimirs are $C_i^k=\sum_{l=1}^n
C_i^{k\,(l)}$, and we can put the sum over
$l$ in front of each subproduct for each of the decompositions. This
would enlarge the number of independent group structures.

Each of the subproducts appearing in a group factor corresponding to a
given number of effective propagators $P$ verifies again
\conditions\ but with a smaller $P$, namely $P^j$. If the number of
subproducts is $s$ it has to be satisfied that,
$$
\sum_{j=1}^s P^j=P.
\eqn\subproducts
$$

It should be noted that if the number $c_i$ in \conditions\ and \ansatz\
were known, our construction would provide a systematic method to
compute the dimension $d_i$. The numbers $c_i$ can probably be
calculated by purely group-theoretical methods. We leave this problem
open for future work.

To clarify how \ansatz\ and the algorithm implicit in the proof works,
some examples are in order. We will  present explicit calculations of
the sequences $\{S_P\}$ for $P\leq 6$, including their corresponding
factors $r_{ij}$, which can be read from \ansatz.  As in the case $P=3$,
we will write integer  sequences following the pattern
$(S(2,1),S(3,1),S(4,1),S(5,1),[S(6,1),S(6,2)],\ldots)$, \ie, in growing
order in $i$ and for a given $i$ in growing order in $k$. The $S(i,k)$
corresponding to a given $i$ will be gathered inside square brackets. As
used above, an  asterisk will indicate that the corresponding group
structure contributes to the framing independent part.

The case $P=4$ is the first one which follows the algorithm. The
sequences $S(i,k)$ are the only possible ones verifying \conditions,\
and can be constructed starting from the data for $P=3$ either by adding
1 to $S(2,1)$ or by adding $-1$ to $S(2,1)$ and 1 to
$S(3,1)$.  In addition, the third possibility pointed out in the proof
of \ansatz\  has to be taken into account because there is a Casimir
corresponding to four effective propagators, $C_4$, and therefore the
sequence
$(0,0,1)$ must be included:
$$
\eqalign{ P=4:& \qquad (4,0,0)\quad \longrightarrow \big(\sum C_2\big)^4
\cr
               & \qquad (2,1,0)\quad \longrightarrow \big(\sum
C_2\big)^2 \sum C_3 \cr
               & \qquad (0,2,0)\quad \longrightarrow \cases{&$\big(\sum
C_3\big)^2 \quad\ast$\cr &$\big(\sum C_2\big) \sum
\big[\big(C_2\big)^{-1}\big(C_3\big)^2\big]$\cr} \cr
               & \qquad (-2,3,0)\quad\longrightarrow\sum
\big[\big(C_2\big)^{-2}\big(C_3\big)^3\big]
\quad\ast\cr
& \qquad (0,0,1)\quad \longrightarrow \sum C_4 \quad\ast
\cr}
$$
Notice that the sequence $(0,2,0)$ contains  more than one group
structure.  This happen after using the generalization of the algorithm
for the semisimple case  which has been described. If the algebra were
simple these multiple structures would reduce to only one, as can be
seen after suppressing the sums over the simple components of the
algebra. To cope with these multiplicities we follow the procedure
outlined in the generalization of the algorithm to semisimple algebras.
First one writes the group structures corresponding to the simple case.
These are:
$
\big(C_2\big)^4,\, \big(C_2\big)^2 C_3,\, \big(C_3\big)^2,\,
\big(C_2\big)^{-2}\big(C_3\big)^3,\, C_4 .
$ Then one considers all possible partitions of the previous factors in
such a way that the subfactors correspond to subdiagrams. These
subfactors correspond to a smaller number of effective propagators, as
explained in \subproducts.\ Once this is done, a sum over the simple
components of the algebra can be put in front of each subfactor. In the
case at hand this can only be done in $\big(C_3\big)^2$ which can also
be written as
$\big[C_2\big]\big[\big(C_3\big)^2\big(C_2\big)^{-1}\big]$. Each of
these two partitions correspond to admissible decompositions in
subdiagrams. The first partition can be represented by two separated
subdiagrams which are three vertices, and the second by one collapsible
propagator besides a subdiagram like the diagram h of Fig. \figvarios.\
Although these two different decompositions lead to the same group
factor in the simple case, the semisimple case distinguishes  them.
Putting the sum over simple components of the algebra in front of each
subfactor we get the two different group structures corresponding to the
sequence $(0,2,0)$. Similar reasonings are followed in the next cases.

The cases $P=5$ and $P=6$ are obtained following the same procedure. We
present here the corresponding results:
$$
\eqalign{ P=5:& \qquad (5,0,0,0)\quad \longrightarrow \big(\sum
C_2\big)^5 \cr
               & \qquad (3,1,0,0)\quad \longrightarrow \big(\sum
C_2\big)^3 \sum C_3 \cr
               & \qquad (1,2,0,0)\quad \longrightarrow
\cases{&$\big(\sum C_2\big)\big(\sum  C_3\big)^2$ \cr & $\big(\sum
C_2\big)^2 \big(\sum \big[\big(C_2\big)^{-1}\big(C_3\big)^2\big]$
\cr}\cr
               & \qquad (1,0,1,0)\quad\longrightarrow \big(\sum C_2\big)
\sum C_4 \cr
               & \qquad (-1,1,1,0)\quad \longrightarrow \sum
\big[\big(C_2\big)^{-1}C_3
C_4\big]\quad\ast\cr
&\qquad (-1,3,0,0)\quad\longrightarrow \cases{&$\sum \big[
\big(C_2\big)^{-1}\big(C_3\big)^2\big]\sum C_3 \quad\ast$\cr
&$\big(\sum  C_2\big)\sum \big[
\big(C_2\big)^{-2}\big(C_3\big)^3\big]$\cr} \cr
               & \qquad (-3,4,0,0)\quad \longrightarrow \sum
\big[\big(C_2\big)^{-3}\big(C_3\big)^4
\big] \quad\ast\cr
               & \qquad (0,0,0,1)\quad \longrightarrow \sum C_5
\quad\ast \cr}
$$
$$\eqalign{ P=6:& \qquad (6,0,0,0,[0,0])\quad \longrightarrow \big(\sum
C_2\big)^6 \cr
               & \qquad (4,1,0,0,[0,0])\quad \longrightarrow \big(\sum
C_2\big)^4 \sum C_3 \cr
               & \qquad (2,2,0,0,[0,0])\quad \longrightarrow
\cases{&$\big(\sum C_2\big)^2\big(\sum C_3\big)^2$\cr &$\big(\sum
C_2\big)^3
\sum\big[\big(C_2\big)^{-1}\big(C_3\big)^2\big]$\cr}
\cr
               & \qquad (2,0,1,0,[0,0])\quad \longrightarrow \big(\sum
C_2\big)^2 \sum C_4 \cr
               & \qquad (1,0,0,1,[0,0])\quad \longrightarrow \big(\sum
C_2\big) \sum C_5 \cr
               & \qquad (0,3,0,0,[0,0])\quad\longrightarrow
\cases{&$\big(\sum C_3\big)^3
\quad\ast$\cr &$\big(\sum C_2\big)\big(\sum
C_3\big)\sum\big[\big(C_2\big)^{-1}\big(C_3\big)^2\big]$\cr &$\big(\sum
C_2\big)^2\sum\big[\big(C_2\big)^{-2}\big(C_3\big)^3\big]$\cr}\cr
               & \qquad (0,1,1,0,[0,0])\quad \longrightarrow
\cases{&$\big(\sum C_3\big) \sum C_4
\quad\ast$\cr &$\big(\sum C_2\big)\sum\big[\big(C_2\big)^{-1}C_3
C_4\big]$\cr}\cr
               & \qquad (-1,1,0,1,[0,0])\quad \longrightarrow
\sum\big[\big(C_2\big)^{-1}C_3 C_5\big]\quad\ast \cr
               & \qquad (-2,2,1,0,[0,0])\quad \longrightarrow
\sum\big[\big(C_2\big)^{-2}\big(C_3\big)^2 C_4\big]\quad\ast \cr
& \qquad (-2,4,0,0,[0,0])\quad \longrightarrow
\cases{&$\big(\sum
C_2\big)\sum\big[\big(C_2\big)^{-3}\big(C_3\big)^4\big]$\cr&$\big(\sum
C_3\big)\sum\big[\big(C_2\big)^{-2}\big(C_3\big)^3\big]
\quad\ast$\cr&$\big(\sum
\big[\big(C_2\big)^{-1}\big(C_3\big)^2\big]\big)^2
\quad\ast$\cr}\cr
               & \qquad (-4,5,0,0,[0,0])\quad \longrightarrow
\sum\big[\big(C_2\big)^{-4}\big(C_3\big)^5\big]\quad\ast\cr
               & \qquad (0,0,0,0,[1,0])\quad \longrightarrow \sum C_6^1
\quad\ast\cr
               & \qquad (0,0,0,0,[0,1])\quad \longrightarrow \sum C_6^2
\quad\ast\cr}
$$

With the help of the results presented for $P=1,\dots,6$, we are in the
position to write
\expansion\ explicitly up to order six:
$$
\eqalign{
\langle W_C^R \rangle = d(R)\Big[1&+ \alpha_{2,1}r_{2,1} x^2 +
\alpha_{3,1}r_{3,1}  x^3 \cr
&+\big(\alpha_{4,1}r_{2,1}^2+\alpha_{4,2}r_{4,2}+\alpha_{4,3}r_{4,3}\big)
x^4 \cr &+
\big(\alpha_{5,1}r_{2,1}r_{3,1}+\alpha_{5,2}r_{5,2}+\alpha_{5,3}r_{5,3}
+\alpha_{5,4}r_{5,4}\big)
x^5 \cr
&+\big(\alpha_{6,1}r_{2,1}^3+\alpha_{6,2}r_{3,1}^2+\alpha_{6,3}r_{2,1}
r_{4,2}+
\alpha_{6,4}r_{2,1}r_{4,3}+\alpha_{6,5}r_{6,5}\cr
&+\alpha_{6,6}r_{6,6}+\alpha_{6,7}r_{6,7}+\alpha_{6,8}r_{6,8}+\alpha_{6,9}
r_{6,9}\big)
x^6 + O(x^7)\Big].\cr}
\eqn\expansiondos
$$ The group factors $r_{ij}$ can be read from the listed results above.
These can be classified in two types, the ones which are not products of
lower order group factors,
$$
\eqalign{ r_{2,1}=&\sum_{k=1}^n C_3^{(k)}, \cr r_{3,1}=&\sum_{k=1}^n
\big(C_3^{(k)}\big)^2
\big(C_2^{(k)}\big)^{-1}, \cr r_{4,2}=&\sum_{k=1}^n \big(C_3^{(k)}\big)^3
\big(C_2^{(k)}\big)^{-2}, \cr r_{4,3}=&\sum_{k=1}^n C_4^{(k)}, \cr
r_{5,2}=&\sum_{k=1}^n
\big(C_3^{(k)}\big)^4 \big(C_2^{(k)}\big)^{-3}, \cr
r_{5,3}=&\sum_{k=1}^n C_4^{(k)} C_3^{(k)}
\big(C_2^{(k)}\big)^{-1}, \cr} \qquad\qquad
\eqalign{ r_{5,4}=&\sum_{k=1}^n C_5^{(k)}, \cr r_{6,5}=&\sum_{k=1}^n
\big(C_3^{(k)}\big)^5
\big(C_2^{(k)}\big)^{-4}, \cr r_{6,6}=&\sum_{k=1}^n C_4^{(k)}
\big(C_3^{(k)}\big)^2
\big(C_2^{(k)}\big)^{-2}, \cr r_{6,7}=&\sum_{k=1}^n C_5^{(k)} C_3^{(k)}
\big(C_2^{(k)}\big)^{-1},
\cr r_{6,8}=&\sum_{k=1}^n C_6^{1\,(k)}, \cr  r_{6,9}=&\sum_{k=1}^n
C_6^{2\,(k)}, \cr}
\eqn\factors
$$  and the ones which are products of lower order ones,
$$
\eqalign{ r_{4,1}=& r_{2,1}^2, \cr
 r_{5,1}=& r_{2,1}r_{3,1}, \cr
 r_{6,1}=& r_{2,1}^3, \cr}
\qquad\qquad
\eqalign{
 r_{6,2}=& r_{3,1}^2, \cr
 r_{6,3}=& r_{2,1}r_{4,2}, \cr
 r_{6,4}=& r_{2,1}r_{4,3}. \cr}
\eqn\losotros
$$ In \FIG\laserres\ Fig. \laserres\ a representative diagram for each
of these group factors has been pictured. These are easily obtained from
\factors\ and \losotros\ after taking into account the diagrams
corresponding to the independent Casimirs drawn in Fig.
\figtresnum.

{}From the results \factors\ and \losotros\ one can read off the values of
the dimensions
$d_i$ for $i=1$ to 6:
$$ d \quad =\quad 0,\quad 1,\quad 1,\quad 3,\quad 4,\quad 9.
\eqn\dimensions
$$ The value $d_1=0$ was implicit in \expansiondos\ and its origin
resides on the absence of the term linear in $x$ in that equation, which
is due to the withdrawal of the framing.

The expansion
\expansion\ verifies a basic property closely related to the
factorization theorem.  If we denote by
$\langle W^{R_k}_C\rangle$ and
$\langle W^R_C
\rangle$ the vacuum expectation values of Wilson lines based on the
algebras $A_k$ and
$A=\oplus_{k=1}^n A_k$, being the representation $R$ a direct product of
representations
$R_k$, it turns out that,
$$
\langle W^R_C \rangle =\prod_{k=1}^n \langle W^{R_k}_C \rangle.
\eqn\factorization
$$ This follows directly from the factorization of both the partition
function and the Wilson line operator. Comparing both sides of
\factorization\  some relations among the geometric coefficients in
their corresponding expansions \expansion\ appear, which can be also
derived after knowing their explicit form by means of the factorization
theorem:
$$
\eqalign{
\alpha_{4,1}=&\half \alpha_{2,1}^2, \cr
\alpha_{5,1}=&\alpha_{2,1}\alpha_{3,1}, \cr
\alpha_{6,1}=&{1\over 6}\alpha_{2,1}^3, \cr}
\qquad\qquad
\eqalign{
\alpha_{6,2}=&\half \alpha_{3,1}^2, \cr
\alpha_{6,3}=&\alpha_{2,1}\alpha_{4,2}, \cr
\alpha_{6,4}=&\alpha_{2,1}\alpha_{4,3}. \cr}
\eqn\factoralpha
$$ These relationships hold also if the Wilson line corresponding to the
knot under consideration is normalized by the Wilson line of the unknot,
although the specific values of the $\alpha_{ij}$ change. This will be
proven in the next section. These properties are explicitly checked in
several examples in the next section.

\endpage

\chapter{The numerical knot invariants}

So far we have been concerned mainly with the group-theoretical aspects
of the perturbative series. Now the numerical coefficients $\alpha_{ij}$
are analyzed. As stated in the introduction they correspond to a series
of line and three-dimensional integrals of certain integrand dictated by
the Feynman rules of Fig. 1. More explicitly, $\alpha_{ij}$ is the sum
of the geometric terms of all diagrams whose group factor contains
$r_{ij}$. The easiest non-trivial example is $\alpha_{2,1}$, which
receives contributions from the diagrams presented in Fig. \casson.\
These  diagrams have the following group factors:
$$
\eqalign{
\tr(T_a T_b T_a T_b)=& \big(\sum C_2\big)^2 d(R)+r_{2,1} d(R), \cr
f_{abc}\tr(T_a T_b T_c)=&-r_{2,1} d(R), \cr}
\eqn\alfadosuno
$$ and therefore, after extracting $x^2$, we can write down a concrete
expression for
$\alpha_{2,1}$,
$$
\eqalign{
\alpha_{2,1}(C)=&{1\over 4{\pi}^2}\oint_C dx_{\mu}\int^x dy_{\nu}\int^y
dz_{\rho}\int^z
dw_{\tau}\epsilon^{\mu\sigma_1\rho}\epsilon^{\nu\sigma_2\tau}
{(x-z)_{\sigma_1}\over
|x-z|^3}{(y-w)_{\sigma_2}\over |y-w|^3}\cr - &{1\over 16{\pi}^3} \oint_C
dx_{\mu}\int^x dy_{\nu}\int^y dz_{\rho}\int_\erretres d^3\omega\Big(
\epsilon^{\mu\rho_1\sigma_1}\epsilon^{\nu\rho_2\sigma_2}
\epsilon^{\rho\rho_3\sigma_3}
\epsilon_{\sigma_1\sigma_2\sigma_3}{(x-w)_{\rho_1}\over |x-w|^3}\cr &
\qquad \qquad \qquad \qquad \qquad \qquad \qquad \qquad
{(y-w)_{\rho_2}\over |y-w|^3}{(z-w)_{\rho_3}\over |z-w|^3}\Big).\cr}
\eqn\alfadosunoint
$$
This quantity was first studied in the context of perturbative
Chern-Simons gauge theory in [\gmmdos]. It turns out that it is related
to the second coefficient in Conway's version of the Alexander
polynomial, and to the Arf and Casson invariants \REF\arfca{L.H.
Kauffman\journal\topo&20(81)101} [\arfca] (see also [\drort]).  Its
value for the unknot, the right-handed trefoil and the figure-eight
knots on the three-manifold $S^3$ are [\gmmdos]:  $-1/6$, $23/6$ and
$-25/6$ respectively.  In [\gmmdos] it was shown explicitly that
$\alpha_{2,1}$ is framing independent. This is guaranteed in our
approach since all framing dependence has been removed from \expansion\
after the identification of diagrams which contribute to framing done in
sect. 2.  Actually, this argument extends to any
${\alpha_{ij}}$, \ie, all these invariants are framing independent. This
invariant is in some sense the paradigm of the finite-type invariants
arisen from perturbative Chern-Simons theory. The next invariant is
$\alpha_{3,1}$. The diagrams needed are $d$,
$e$, $f$ and $h$ of Fig. \figvarios\ whose group factors are
$$
\eqalign{
\tr(T_a T_b T_a T_c T_b T_c)=&\big(\sum C_2\big)^3d(R)+2\big(\sum
C_2\big) r_{2,1}d(R) +r_{3,1}d(R),\cr \tr(T_a T_b T_c T_a T_b
T_c)=&\big(\sum C_2\big)^3d(R)+3\big(\sum C_2\big) r_{2,1}d(R)
+2r_{3,1}d(R),\cr f_{abc}\tr(T_a T_b T_d T_c T_d)=&\big(\sum
C_2\big)r_{2,1}d(R)+r_{3,1}d(R),\cr f_{abr}f_{rcd}\tr(T_a T_b T_c
T_d)=&r_{3,1}d(R).\cr}
\eqn\alfatresuno
$$ Taking into account that there are 3 diagrams of type $d$, 1 of type
$e$, 5 of type $f$ and 2 of type $h$ in the perturbative expansion, the
integral representation of
$\alpha_{3,1}$ reads
$$
\eqalign{
\alpha_{3,1}(C)=&{3\over 8\pi^3}\oint_C dx_{\mu}\int^x dy_{\nu}\int^y
dt_{\rho}\int^t dz_{\tau}\int^z dv_{\eta} \int^v dw_{\zeta}
\epsilon^{\mu\sigma_1\tau}\epsilon^{\nu\sigma_2\zeta}
\epsilon^{\rho\sigma_3\eta}\cr
& {(x-z)_{\sigma_1}\over |x-z|^3}{(y-w)_{\sigma_2}\over
|y-w|^3}{(t-v)_{\sigma_3}\over |t-v|^3}\cr+ & {1\over 4\pi^3}\oint_C
dx_{\mu}\int^x dy_{\nu}\int^y dt_{\rho}\int^t dz_{\tau}\int^z dv_{\eta}
\int^v dw_{\zeta}
\epsilon^{\mu\sigma_1\tau}\epsilon^{\nu\sigma_2\eta}
\epsilon^{\rho\sigma_3\zeta}\cr
& {(x-z)_{\sigma_1}\over |x-z|^3}{(y-v)_{\sigma_2}\over
|y-v|^3}{(t-w)_{\sigma_3}\over |t-w|^3}\cr+ & {5\over 32\pi^4}\oint_C
dx_{\mu}\int^x dy_{\nu}\int^y dt_{\rho}\int^t dz_{\tau}\int^z
dv_{\eta}\int_\erretres d^3\omega
\epsilon^{\nu\sigma\eta}\epsilon_{\alpha\beta\gamma}
\epsilon^{\mu\sigma_1\alpha}
\epsilon^{\rho\sigma_2\beta}\epsilon^{\tau\sigma_3\gamma}\cr
&{(y-v)_{\sigma}\over |y-w|^3}{(x-\omega)_{\sigma_1}\over
|x-\omega|^3}{(t-\omega)_{\sigma_2}\over
|t-\omega|^3}{(z-\omega)_{\sigma_3}\over |z-\omega|^3}\cr + & {1\over
64\pi^5}\oint_C dx_{\mu}\int^x dy_{\nu}\int^y dt_{\rho}\int^t
dz_{\tau}\int_\erretres d^3\omega_1\int_\erretres d^3\omega_2
\epsilon_{\alpha\beta\gamma}\epsilon_{\eta\xi\zeta}
\epsilon^{\mu\sigma_1\alpha}\cr
&\epsilon^{\nu\sigma_2\beta}\epsilon^{\gamma\sigma_3\zeta}
\epsilon^{\rho\sigma_4\eta}\epsilon^{\tau\sigma_5\xi}
{(x-\omega_1)_{\sigma_1}\over
|x-\omega_1|^3}{(y-\omega_1)_{\sigma_2}\over
|y-\omega_1|^3}{(\omega_1-\omega_2)_{\sigma_3}\over
|\omega_1-\omega_2|^3}{(t-\omega_2)_{\sigma_4}\over
|t-\omega_2|^3}{(z-\omega_2)_{\sigma_5}\over |z-\omega_2|^3}\cr}
\eqn\alfatresunoint
$$

An important property of the geometrical factors $\alpha_{ij}$ is their
behavior under changes of orientation in  the manifold. Notice that
while $\alpha_{2,1}$ possesses a product of an even number of
three-dimensional totally antisymmetric tensors in all its terms,
$\alpha_{3,1}$ has a product of an odd number. Thus, under a change of
orientation
$\alpha_{2,1}$ is even and $\alpha_{3,1}$ is odd. From the Feynman rules
of the theory follows that, in general, for even $i$ the factors
$\alpha_{ij}$ are even under a change of orientation while for odd $i$
those factors are odd. This implies that a knot $K$ and its mirror image
$\tilde K$ have geometrical factors such that
$$
\eqalign{
\alpha_{ij}(K) & = \alpha_{ij}(\tilde K),  \quad {\hbox{\rm if}} \quad i
\quad {\hbox{\rm is even}}, \cr
\alpha_{ij}(K) & = - \alpha_{ij}(\tilde K),  \quad {\hbox{\rm if}} \quad
i \quad {\hbox{\rm is odd}}.\cr}
\eqn\evenodd
$$ In particular, for amphicheiral knots ($K \sim \tilde K$),
$\alpha_{ij}(K)=0$ for
$i$ odd. Looking back at the expansion \expansion\ one observes that
these results are in agreement with the fact that for quantum group knot
invariants, their value for knots related by a change of orientation in
the manifold are the same once the replacement $q \rightarrow q^{-1}$ is
performed. Since
$q=\ex^x$, this is equivalent to carry out the  change $x \rightarrow
-x$, which, using
\expansion\ and the fact that the $r_{ij}$ are knot independent implies
\evenodd.

It is possible to continue the procedure described above and give the
expressions which correspond to higher coefficients $\alpha_{ij}$. One
simply has to draw all diagrams corresponding to the given order,
compute their group factors in the way explained above, gather the
framing independent contributions and display an integral after using
the Feynman rules. Nevertheless the resulting expressions are somewhat
unwieldy and not too illuminating. We will instead study the properties
of these knot invariants.

The  knot invariants  $\alpha_{ij}$ can be written in many ways because
their defining expansion \expansion\ is subject to two different types
of normalizations. On the one hand, the vacuum expectation value
$\langle W^R_C \rangle$ could be normalized differently. For example,
the choice made in \expansion\ is such that it does not have value one
for the unknot. Dividing  $\langle W^R_C \rangle$ by the corresponding
quantity for the unknot will shift the values of the $\alpha_{ij}$. On
the other hand, the group factors depend on the group theoretical
conventions, in particular the normalization of the generators of the
semisimple group. Since
$\langle W^R_C \rangle$ is independent on how those generators are
normalized, the
$\alpha_{ij}$ must be different for different normalizations. In more
explicit terms, the integral expressions obtained for $\alpha_{2,1}$ and
for $\alpha_{3,1}$ in \alfadosunoint\ and \alfatresunoint\ would contain
different global factors. To make the knot invariants $\alpha_{ij}$
universal we will first redefine them dividing by the unknot. This will
fix the additive arbitrariness of the
$\alpha_{ij}$ and  will impose the property that all these invariants
vanish for the unknot. Second, we will fix the multiplicative
arbitrariness of the invariants
$\alpha_{ij}$ by taking the simplest non-trivial knot, the trefoil, and
fixing the values of the invariants to some selected integers. This can
be done if all the invariants do not vanish for the trefoil. This holds
up to order six and we will assume that it holds in general. As we will
see in the next section the choice made supports the conjecture that it
is possible to find a normalization were all the invariant quantities
are integers. This is a highly non-trivial feature looking
at their integral representations as the ones in \alfadosunoint\ and
\alfatresunoint. In the next section we will present all these facts
explicitly up to order six for all prime knots up to six crossings.

Let us denote the unknot by $U$ and let us consider its expansion
\expansion,
$$
\langle W^R_U \rangle = d(R)
\sum_{i=0}^\infty \sum_{j=1}^{d_i}\alpha_{ij}(U)r_{ij}(R) x^i.
\eqn\expansionunknot
$$ Let us now consider un arbitrary knot $K$. We define the new knot
invariants
$\tilde\alpha_{ij}(K)$ normalizing by the expression for the unknot:
$$ {\langle W^R_K \rangle\over\langle W^R_U \rangle}= {\sum_{i=0}^\infty
\sum_{j=1}^{d_i}\alpha_{ij}(K)r_{ij}(R) x^i \over
 \sum_{i=0}^\infty \sum_{j=1}^{d_i}\alpha_{ij}(U)r_{ij}(R) x^i} =
\sum_{i=0}^\infty \sum_{j=1}^{d_i}\tilde\alpha_{ij}(K)r_{ij}(R) x^i.
\eqn\alfatilde
$$ Similarly to the case of \expansion, one has $\tilde\alpha_{0,1}=1$.
Notice that for the unknot, $U$, one has $\tilde\alpha_{ij}(U)=0$,
$\forall\, i,j$, such that $i\not= 0$. From the values given above for
$\alpha_{2,1}$ for the unknot, the right-handed trefoil and the
figure-eight knot ($-1/6$, $23/6$ and $-25/6$ respectively), one easily
obtain the value of $\tilde\alpha_{2,1}$ for the right-handed trefoil
and the figure-eight knot: $23/6 + 1/6= 4$ and $-25/6 +1/6=-4$,
respectively. It is clear from
\alfatilde\ and the fact that the unknot is amphicheiral that the
properties \evenodd\ are also satisfied by the $\tilde\alpha_{ij}$.

The new quantities $\tilde\alpha_{ij}(K)$ also satisfy relations as the
ones in
\factoralpha. To prove this notice that relation \factorization\ holds
for any knot, in particular for the unknot. This implies,
$$ {\langle W^R_C \rangle \over \langle W^R_U \rangle} = {\prod_{k=1}^n
\langle W^{R_k}_C \rangle \over \prod_{k=1}^n \langle W^{R_k}_U \rangle}
=
\prod_{k=1}^n {\langle W^{R_k}_C \rangle \over  \langle W^{R_k}_U
\rangle},
\eqn\melon
$$ which, similarly to the case \factoralpha\ leads to,
$$
\eqalign{
\tilde\alpha_{4,1}=&\half \tilde\alpha_{2,1}^2, \cr
\tilde\alpha_{5,1}=&\tilde\alpha_{2,1}\tilde\alpha_{3,1}, \cr
\tilde\alpha_{6,1}=&{1\over 6}\tilde\alpha_{2,1}^3, \cr}
\qquad\qquad
\eqalign{
\tilde\alpha_{6,2}=&\half \tilde\alpha_{3,1}^2, \cr
\tilde\alpha_{6,3}=&\tilde\alpha_{2,1}\tilde\alpha_{4,2}, \cr
\tilde\alpha_{6,4}=&\tilde\alpha_{2,1}\tilde\alpha_{4,3}. \cr}
\eqn\uva
$$

The knot invariants $\tilde\alpha_{ij}$, besides being topological
invariants and framing independent,  are also  knot invariants of finite
type in the sense of Vassiliev. This follows from the results  of
[\birlin], where the  authors showed that the
$i^{\rm{th}}$ coefficient of the expansion in $x$ of the HOMFLY and
Kauffman polynomials of an arbitrary knot $K$, $H_{N,q}(K)$ and
$R_{N,q}(K)$, after taking $q=\ex^x$, is a Vassiliev invariant of order
$i$.  The theorem was extended for an arbitrary quantum group invariant
in [\lin]. It is also presented in [\drort] and, in full generality, in
\REF\birman{J.S. Birman
\journal\bams&28(93)253} [\birman]. We simply extend this result to the
different coefficients
$\tilde\alpha_{ij}$ which contribute to the
$i^{\rm{th}}$ order in the expansion. The idea is that at a given order
the different structures
$r_{ij}$ are independent and therefore all the
$\tilde\alpha_{ij}$ are independent Vassiliev invariants of order $i$.

To be  self-contained, we review very  briefly the axiomatic approach
to  Vassiliev invariants proposed in [\birlin] and [\birman], where a
thorough treatment of the subject can be found. A
$j$-singular knot is a knot which has
$j$ transversal self-intersections. This object is denoted by $K^j$. The
self-intersection can be made an undercrossing or an overcrossing, which
are called the resolutions of the self-intersection. Given a knot
invariant
$V(K)$ it can be extended to be an invariant of $j$-singular knots by
means of the  prescription presented in
\FIG\crossings\ Fig. \crossings.

The formula presented in Fig. \crossings\ is the first axiom of Birman
and Lin. If we denote by
$K_+^j$ a $j$-singular knot with an overcrossing at a given point and by
$K_-^j$ the same with an undercrossing instead of the overcrossing, this
axiom can be written as a crossing-change  formula,
$$ V(K^j)=V(K_+^{j-1})-V(K_-^{j-1}).
\eqn\axiomone
$$ The second axiom states that the Vassiliev invariants  vanish on
$j$-singular knots for $j$ high enough. In other words:
$$
\exists i \in Z^+ \qquad{\rm{such\,\, that}}\qquad V(K^j)=0
\qquad{\rm{if}} \qquad j>i.
\eqn\axiomtwo
$$ The smallest such $i$ is called the order (or type) of $V$. A
Vassiliev invariant of order
$i$ will be denoted by $V_i$. Besides these two axioms, some initial
data are needed. Let us denote by
$U$ the unknot. Then one requires that all Vassiliev invariants of $U$
vanish,
$$ V_i(U)=0, \qquad \forall i\in Z^+.
\eqn\axiomfour
$$ A singular point $p$ is called nugatory if its two resolutions
define the same knot. Let us denote by $K_p^j$ a singular knot which
includes a nugatory point $p$. If we are  to obtain knot invariants, the
following axiom has to be satisfied:
$$ V_i(K_p^j)=0, \qquad{\rm{if\,\, {\it{p}} \,\,is\,\, nugatory}}.
\eqn\axiomfive
$$ A further set of initial data is needed to begin with the calculation
of the invariants. This corresponds to a set of given $V_i(K^j)$ for
some selected singular knots, presented in the form of a table, called
the actuality table. Of course these numbers are not arbitrary and
have to satisfy some rules in order to yield consistent values for the
numerical invariants. These consistency conditions are a system of
linear equations, where the unknowns are the numbers present in the
actuality table. Birman and Lin [\birlin,\birman] proved that the
expansion of any quantum group invariant associated to a knot $K$ yields
consistent values for these numerical knot invariants, and therefore are
Vassiliev invariants. We will use this result to prove that the knot
invariants $\tilde\alpha_{ij}$ in \alfatilde\ are Vassiliev invariants.

Let us consider a knot $K$. According to Birman and Lin [\birlin,\lin],
for any semisimple group, one can assert that the contribution  at each
order in $x$ in the expansion \expansion,
$$ V_i(K)=\sum_{k=1}^{d_i} \tilde\alpha_{ik}(K) r_{ik},
\eqn\vastot
$$ is a Vassiliev invariant of order $i$. This means that these $V_i$
satisfy the axioms given above. In other words, if one defines
invariants for $j$-singular knots from \vastot\ using \axiomone, one
finds that \axiomtwo, \axiomfour\ and \axiomfive\ hold, and that a
consistent actuality table is obtained. Following the same mechanism we
define the $\tilde\alpha_{ik}(K)$ for
$j$-singular knots using
\vastot\ and
\axiomone. Indeed, in writing \axiomone, one finds,
$$
 \sum_{k=1}^{d_i} \tilde\alpha_{ik}(K^j) r_{ik} =
 \sum_{k=1}^{d_i} \tilde\alpha_{ik}(K^{j-1}_+) r_{ik} -\sum_{k=1}^{d_i}
\tilde\alpha_{ik}(K^{j-1}_-) r_{ik},
\eqn\limon
$$ and then, from the independence of the group factors $r_{ik}$ follows
that,
$$
\tilde\alpha_{ik}(K^j) = \tilde\alpha_{ik}(K^{j-1}_+) -
\tilde\alpha_{ik}(K^{j-1}_-).
\eqn\sandia
$$ To prove that the quantities $\tilde\alpha_{ik}(K^j)$ defined in this
way satisfy \axiomtwo,
\axiomfour\ and \axiomfive\ we must take into account that, again,
making use of the theorem by Birman and Lin,  $V_i(K^j)=\sum_{k=1}^{d_i}
\tilde\alpha_{ik}(K^j) r_{ik}$ do satisfy these axioms.  Writting out
the corresponding expressions in terms of this sum and making use of the
independence of the group factors follows that  \axiomtwo,
\axiomfour\ and \axiomfive\ are also verified by the quantities
$\tilde\alpha_{ik}(K^j)$. Similarly, one concludes that given a type $i$
there is a consistent actuality table for each
$k$, \ie, the $\tilde\alpha_{ik}$ generate an actuality table once their
value for $j$-singular knots is defined through \sandia. Therefore, the
geometrical factors associated to knots in
\alfatilde\ and  their extension to $j$-singular knots done in \sandia\
are invariants of finite type or Vassiliev invariants. Notice that these
invariants generate $d_i$ actuality tables for each $i$. The actuality
table that one would generate following Birman and Lin for a given
quantum group invariant would be a special linear combination of these
$d_i$ actuality tables.

Vassiliev invariants form an algebra, not just a sequence of vector
spaces.  Products of Vassiliev invariants of types $i$ and $j$ lead to
Vassiliev invariants of type $ij$. This structure is also manifest in
our knot invariants due to relations
\uva. These follow from the factorization property of Wilson lines for
semisimple groups  \factorization. Invariants can in this way be
classified in two types: simple invariants as the ones which are not
product of lower type invariants, and compound invariants which are the
rest. Taking into account \uva, $\tilde\alpha_{2,1}$,
$\tilde\alpha_{3,1}$, $\tilde\alpha_{4,2}$, $\tilde\alpha_{4,3}$,
$\tilde\alpha_{5,2}$, $\tilde\alpha_{5,3}$, $\tilde\alpha_{5,4}$,
$\tilde\alpha_{6,5}$, $\tilde\alpha_{6,6}$, $\tilde\alpha_{6,7}$,
$\tilde\alpha_{6,8}$ and $\tilde\alpha_{6,9}$ are simple invariants. The
rest are compound invariants. An important quantity is the number of
simple invariants at each order or type. We will denote it by $\hat d_i$.
For $i=1,\dots,6$ one finds,
$$
\hat d \quad =\quad 0,\quad 1,\quad 1,\quad 2,\quad 3,\quad 5.
\eqn\dimensions
$$

One of the  most important aspects of this work is that it provides a
geometrical interpretation of  Vassiliev invariants in the sense that
they are written as integrations of the type \alfadosunoint\ and
\alfatresunoint\ and their generalizations.  It would be interesting to
study the relation between this  integral representation and the one by
Kontsevich \REF\konse{M. Kontsevich, ``Graphs, homotopical algebra and
low-dimensional topology", preprint, 1992} [\konse]. Our approach is
intrinsic to three dimensions and can be easily generalized to arbitrary
three-manifold. In contrast,  Kontsevitch's representation is basically
two-dimensional (in the sense that the three-manifold is considered as a
product ${\bf R}\times {\bf C}$) and therefore it is not obvious how to
extend it to more general situations. However, for the case in which
Kontsevich representation is defined, they should be related. This is
supported by recent work \REF\mateos{J. Mateos, ``Ghosts, knots and
Kontsevich integrals",  DAMTP/93-05 preprint, 1993} [\mateos] showing
that Chern-Simons gauge theory in the Hamiltonian formalism leads to
Kontsevich representation for Vassiliev invariants.

It is important at this point  to discuss  the relation between this
work and the one by Bar-Natan  \REF\drortopo{D. Bar-Natan, ``On the
Vassiliev knot invariants", Harvard University preprint, 1993} in
[\drortopo]. In [\drortopo] it is proved that the group factors of
 all Feynman diagrams with no collapsible propagators which contribute
to a given order in $g^2$ (or $x$) can be regarded as Vassiliev
invariants.  These Vassiliev invariants have an entirely different
origin than the knot invariants
$\tilde\alpha_{ik}$. Bar-Natan's approach has two steps. First the
observation that one can associate  Feynman-like diagrams with no
collapsible propagators to $j$-singular knots. Second that assigning the
group factors of these diagrams to $j$-singular knots one constructs a
set of rational numbers (weight system) that satisfies the axioms by
Birman and Lin.   Clearly, this observation  is orthogonal to our
results. It is important, however, to notice that the dimension of the
space of Vassiliev invariants in [\drortopo] equals the number of
independent group structures and therefore must have the same values as
our $d_i$. Comparing the results in [\drortopo] with the values for
$d_i$ presented  in \dimensions\ one finds complete agreement up to
$i=6$. We think that the  calculation of these dimensions, in general,
is more tractable in our approach. Recall that, as stated in the
previous section, after  equation \ansatz\ and its generalization for
semisimple groups, the problem to compute $d_i$ is reduced to the
problem of finding $c_i$, or number of independent Casimirs of order
$i$. The study of this issue is left for future work.

Taking into account  the work [\drortopo] there appears a very appealing
situation. The coefficient of $x^i$ in the
expansion of any quantum group invariant can be regarded as the inner
product of two vectors of dimension $d_i$ (the $\alpha_{ik}$ and the
$r_{ik}$, $k=1,\cdots,d_i$). Each of these two vectors is made out of
Vassiliev invariants though each one has a different interpretation:
while the $\alpha_{ik}$ are associated to the non-singular knot under
study, the $r_{ik}$ are in correspondence with a very precise set of
$j$-singular knots, and associated to the group and representation under
consideration.

\endpage

\chapter{Numerical knot invariants for all prime  knots up to six
crossings}

In this section we will present the calculation of the knot invariants
$\tilde\alpha_{ij}$ for all prime knots up to six crossings up to order
six. Then we will show evidence supporting that there exist a
normalization such that these knot invariants are integer-valued.

In order to compute the knot invariants $\tilde\alpha_{ij}(K)$ one could
try to evaluate their integral expressions. This is certainly a long and
tedious way to proceed. There is a faster way to carry out their
computation using known information on the left hand side of \alfatilde.
Indeed, the knot invariant
$\langle W^R_K \rangle / \langle W^R_U \rangle$ is known for a variety
of groups and representations for many knots on the three-manifold
$S^3$.  Taking its value for different cases one generates systems of
linear equation where the unknowns are the $\tilde\alpha_{ij}(K)$. Recall
that while the $r_{ij}(R)$ are group and representation dependent, they
are knot independent. All dependence on $K$ is contained in
$\tilde\alpha_{ij}(K)$ which, on the other hand, are independent of the
group and the representation. Up to order
$i=6$, which is the situation analyzed in this section,  it is enough to
consider the following cases: $SU(2)$ in an arbitrary representation of
spin $j$ (Jones and Akutsu-Wadati polynomials
[\jones,\jonesAM,\aw,\poli,\kguno]), $SU(N)$ in the fundamental
representation  $f$ (HOMFLY polynomial [\homflyp,\jonesAM]), $SO(N)$ in
the fundamental representation (Kauffman polynomial [\kauf]), and
$SU(2)\times SU(N)$ in representations of the form $(j,f)$, \ie, a
representation of spin
$j$ in the  subgroup $SU(2)$,  and the fundamental in the  subgroup
$SU(N)$. These invariants are known and can be collected from the
literature. They are listed in Appendix C.

The structure of the computation to be carried out is the following.
Once the polynomial invariant corresponding to the left hand side of
\alfatilde\ is collected one replaces its variable $q$ by $\ex^x$ and
expands in powers of $x$. For the case considered in this section one
needs just the expansion up to order six. The coefficients of $x^i$ are
either polynomials  in $N$, polynomials in $j$, or polynomials in $N$
and $j$. On the other hand, on the right hand side of \alfatilde\ the
group factors are the ones in \factors\ and \losotros, which can be
written explicitly using the values of the corresponding Casimirs, which
are listed in Appendix B. Again, one observes quickly that the group
factors are polynomials in $N$, $j$, or both, $N$ and $j$. Both sides of
\alfatilde\ must then be compared. This leads to series of linear
equations which must be satisfied by the  $\tilde\alpha_{ij}(K)$. It
turns out that one encounters 5 equations for $\tilde\alpha_{2,1}(K)$, 5
equations for
$\tilde\alpha_{3,1}(K)$, 12 equations for $\tilde\alpha_{4,1}(K)$,
$\tilde\alpha_{4,2}(K)$ and $\tilde\alpha_{4,3}(K)$, 15 equations for
$\tilde\alpha_{5,1}(K)$, $\dots$, $\tilde\alpha_{5,4}(K)$, and 20
equations for
$\tilde\alpha_{6,1}(K)$, $\dots$, $\tilde\alpha_{6,9}(K)$. Those
equations determine uniquely all the $\tilde\alpha_{ij}(K)$ up to order
six.

The values of the $\tilde\alpha_{ij}(K)$ obtained in this way are, in
general, rational numbers. For the trefoil, which will be labeled in the
standard form $3_1$
\REF\rolfsen{D. Rolfsen, {\it Knots and links}, 1976, Berkeley, Calif.:
Publish or Perish Press} [\rolfsen], one finds,
$$
\eqalign{
\tilde\alpha_{2,1}(3_1) &=4    ,\cr
\tilde\alpha_{3,1}(3_1) &=8    ,\cr
\tilde\alpha_{4,1}(3_1) &=8    ,\cr
\tilde\alpha_{4,2}(3_1) &= {62\over 3}   ,\cr
\tilde\alpha_{4,3}(3_1) &={10\over 3}    ,\cr
\tilde\alpha_{5,1}(3_1) &=32    ,\cr
\tilde\alpha_{5,2}(3_1) &={176\over  3}    ,\cr
\tilde\alpha_{5,3}(3_1) &={32\over 3}    ,\cr
\tilde\alpha_{5,4}(3_1) &=8    ,\cr}
\qquad\qquad
\eqalign{
\tilde\alpha_{6,1}(3_1) &={32\over 3}    ,\cr
\tilde\alpha_{6,2}(3_1) &=32    ,\cr
\tilde\alpha_{6,3}(3_1) &={248\over 3}   ,\cr
\tilde\alpha_{6,4}(3_1) &={40\over 3}   ,\cr
\tilde\alpha_{6,5}(3_1) &={5071\over 30}    ,\cr
\tilde\alpha_{6,6}(3_1) &={116\over 30}    ,\cr
\tilde\alpha_{6,7}(3_1) &={3062\over 45}    ,\cr
\tilde\alpha_{6,8}(3_1) &={17\over 18}    ,\cr
\tilde\alpha_{6,9}(3_1) &={271\over 30}    .\cr}
\eqn\trebolalfas
$$ Notice that these quantities satisfy the relations predicted in \uva.
We will not present the values of the $\tilde\alpha_{ij}$ for other
knots since we are going first to normalize them properly.

As discussed in the previous section, these $\tilde\alpha_{ij}(K)$ are
not universal in the sense that they depend on the group theoretical
conventions used. It should be desirable to redefine them in such a way
that they do not depend on those conventions. The simplest way to
proceed would be to decide that these knot invariants take the value  1
for some knot in which none of them vanish. We will assume that this
non-vanishing feature occurs for the simplest knot, the trefoil. It is
certainly true up to order six as can be seen in
\trebolalfas.   As we will show below, there is another normalization
possibility, which is the one that we will finally take, in which all
invariants up to order six seem to be integer-valued. Choosing the
values for  $\tilde\alpha_{ij}(3_1)$ as some selected integers, it turns
out that the resulting invariants are integers. Our computations show
that this  happens to all prime knots up to six crossings. This leads
us to conjecture that a similar picture holds for all orders and all
knots.

Let  us first redefine the universal knot invariants. We will denote
them by
$\tilde\beta_{ij}(K)$ as,
$$
\tilde\beta_{ij}(K) = {\tilde\alpha_{ij}(K)\over\tilde\alpha_{ij}(3_1)}.
\eqn\betatilde
$$ These invariants  are well defined if the quantities
$\tilde\alpha_{ij}(3_1)$ do not vanish. We will assume this holds. On
the other hand, it is clear from
\betatilde\ that the invariants $\tilde\beta_{ij}(K)$ are independent of
the normalization chosen for the  Lie algebra generators. There is
certainly a dependence on the basis for group factors which has been
chosen but this dependence is intrinsic to their definition. For the
right-handed trefoil, $\tilde\beta_{ij}(3_1)=1$. In this normalization
the algebra \uva\ simplifies. Instead of having numerical factors as the
ones in \uva\ it turns out that just the product of two invariants, one
of order $i$ and another of order $j$, leads to the corresponding
invariant of order $ij$.

Actually, it seems that there  is a refined choice of the universal knot
invariants which makes them integer-valued. This choice can only be
observed after computing $\tilde\beta_{ij}(K)$ for many knots. We have
carried out their calculation for all prime knots up to six crossings.
The selected normalization up to order six is,
$$
\eqalign{
\beta_{2,1}(K) &=    {\tilde\alpha_{2,1}(K) \over
\tilde\alpha_{2,1}(3_1)},\cr
\beta_{3,1}(K) &=    {\tilde\alpha_{3,1}(K) \over
\tilde\alpha_{3,1}(3_1)},\cr
\beta_{4,2}(K) &=  31{\tilde\alpha_{4,2}(K) \over
\tilde\alpha_{4,2}(3_1)},\cr
\beta_{4,3}(K) &=   5{\tilde\alpha_{4,3}(K) \over
\tilde\alpha_{4,3}(3_1)},\cr
\beta_{5,2}(K) &=  11{\tilde\alpha_{5,2}(K) \over
\tilde\alpha_{5,2}(3_1)},\cr
\beta_{5,3}(K) &=    {\tilde\alpha_{5,3}(K) \over
\tilde\alpha_{5,3}(3_1)},\cr}
\qquad\qquad
\eqalign{
\beta_{5,4}(K) &=    {\tilde\alpha_{5,4}(K) \over
\tilde\alpha_{5,4}(3_1)},\cr
\beta_{6,5}(K) &=5071{\tilde\alpha_{6,5}(K) \over
\tilde\alpha_{6,5}(3_1)},\cr
\beta_{6,6}(K) &=  29{\tilde\alpha_{6,6}(K) \over
\tilde\alpha_{6,6}(3_1)},\cr
\beta_{6,7}(K) &=1531{\tilde\alpha_{6,7}(K) \over
\tilde\alpha_{6,7}(3_1)},\cr
\beta_{6,8}(K) &=  17{\tilde\alpha_{6,8}(K) \over
\tilde\alpha_{6,8}(3_1)},\cr
\beta_{6,9}(K) &= 271{\tilde\alpha_{6,9}(K) \over
\tilde\alpha_{6,9}(3_1)}.\cr}
\eqn\trebolbetas
$$ Notice that we have written only the simple knot invariants. The
compound knot invariants are defined in such a way that they are
products of simple ones:
$$
\eqalign{
\beta_{4,1}=&\beta_{2,1}^2, \cr
\beta_{5,1}=&\beta_{2,1}\beta_{3,1}, \cr
\beta_{6,1}=&\beta_{2,1}^3, \cr}
\qquad\qquad
\eqalign{
\beta_{6,2}=&\beta_{3,1}^2, \cr
\beta_{6,3}=&\beta_{2,1}\beta_{4,2}, \cr
\beta_{6,4}=&\beta_{2,1}\beta_{4,3}. \cr}
\eqn\uvados
$$

The explicit expressions of these universal simple knot invariants  for
all prime knots up to six crossings are presented in Table I.  Knots
have been labeled in their standard form  [\rolfsen].

\vskip1cm
\vbox{
\vbox{\tabskip=0pt
\offinterlineskip
\def\tablerule{\noalign{\hrule}}
\halign { \vrule# \tabskip=0.3em plus2em minus.5em
         & \hfil#\hfil & \vrule#
         & \hfil#\hfil & \vrule# & \hfil#\hfil & \vrule#
         & \hfil#\hfil & \vrule# & \hfil#\hfil & \vrule#
         & \hfil#\hfil & \vrule# & \hfil#\hfil & \vrule#
         & \hfil#\hfil & \vrule# & \hfil#\hfil & \vrule#
         & \hfil#\hfil & \vrule# & \hfil#\hfil & \vrule#
         & \hfil#\hfil & \vrule# & \hfil#\hfil & \vrule# \tabskip=0pt \cr
\tablerule
 height2pt &\omit && \omit && \omit && \omit && \omit && \omit && \omit
&&
            \omit && \omit && \omit && \omit && \omit && \omit& \cr
  & knot && $\beta_{2,1}$  && $\beta_{3,1}$ && $\beta_{4,2}$ &&
$\beta_{4,3}$
         && $\beta_{5,2}$  && $\beta_{5,3}$ && $\beta_{5,4}$ &&
$\beta_{6,5}$
         && $\beta_{6,6}$ &&  $\beta_{6,7}$ && $\beta_{6,8}$ &&
$\beta_{6,9}$ & \cr
 height2pt &\omit && \omit && \omit && \omit && \omit && \omit && \omit
&&
            \omit && \omit&& \omit && \omit && \omit && \omit& \cr
\tablerule
 height2pt &\omit && \omit && \omit && \omit && \omit && \omit && \omit
&&
            \omit && \omit && \omit && \omit && \omit && \omit& \cr
  & $0_1$ && $0$  && $0$ && $0$ && $0$
         && $0$  && $0$ && $0$ && $0$
         && $0$ &&  $0$ && $0$ && $0$ & \cr
 height2pt &\omit && \omit && \omit && \omit && \omit && \omit && \omit
&&
            \omit && \omit&& \omit && \omit && \omit && \omit& \cr
\tablerule
 height2pt &\omit && \omit && \omit && \omit && \omit && \omit && \omit
&&
            \omit && \omit && \omit && \omit && \omit && \omit& \cr
  & $3_1$ && $1$  && $1$ && $31$ && $5$
         && $11$  && $1$ && $1$ && $5071$
         && $29$ &&  $1531$ && $17$ && $271$ & \cr
 height2pt &\omit && \omit && \omit && \omit && \omit && \omit && \omit
&&
            \omit && \omit&& \omit && \omit && \omit && \omit& \cr
\tablerule
 height2pt &\omit && \omit && \omit && \omit && \omit && \omit && \omit
&&
            \omit && \omit && \omit && \omit && \omit && \omit& \cr
  & $4_1$ && $-1$  && $0$ && $17$ && $7$
         && $0$  && $0$ && $0$ && $-1231$
         && $71$ &&  $-871$ && $79$ && $-271$ & \cr
 height2pt &\omit && \omit && \omit && \omit && \omit && \omit && \omit
&&
            \omit && \omit&& \omit && \omit && \omit && \omit& \cr
\tablerule
 height2pt &\omit && \omit && \omit && \omit && \omit && \omit && \omit
&&
            \omit && \omit && \omit && \omit && \omit && \omit& \cr
  & $5_1$ && $3$  && $5$ && $261$ && $39$
         && $157$  && $14$ && $13$ && $123453$
         && $1247$ &&  $34353$ && $387$ && $5853$ & \cr
 height2pt &\omit && \omit && \omit && \omit && \omit && \omit && \omit
&&
            \omit && \omit&& \omit && \omit && \omit && \omit& \cr
\tablerule
 height2pt &\omit && \omit && \omit && \omit && \omit && \omit && \omit
&&
            \omit && \omit && \omit && \omit && \omit && \omit& \cr
  & $5_2$ && $2$  && $3$ && $134$ && $22$
         && $69$  && $6$ && $7$ && $45902$
         && $-42$ &&  $14882$ && $274$ && $2702$ & \cr
 height2pt &\omit && \omit && \omit && \omit && \omit && \omit && \omit
&&
            \omit && \omit&& \omit && \omit && \omit && \omit& \cr
\tablerule
 height2pt &\omit && \omit && \omit && \omit && \omit && \omit && \omit
&&
            \omit && \omit && \omit && \omit && \omit && \omit& \cr
  & $6_1$ && $-2$  && $1$ && $58$ && $26$
         && $-19$  && $-2$ && $-3$ && $-5582$
         && $442$ &&  $-5042$ && $686$ && $-1742$ & \cr
 height2pt &\omit && \omit && \omit && \omit && \omit && \omit && \omit
&&
            \omit && \omit&& \omit && \omit && \omit && \omit& \cr
\tablerule
 height2pt &\omit && \omit && \omit && \omit && \omit && \omit && \omit
&&
            \omit && \omit && \omit && \omit && \omit && \omit& \cr
  & $6_2$ && $-1$  && $1$ && $17$ && $19$
         && $-13$  && $-2$ && $-3$ && $2129$
         && $331$ &&  $-931$ && $463$ && $-751$ & \cr
 height2pt &\omit && \omit && \omit && \omit && \omit && \omit && \omit
&&
            \omit && \omit&& \omit && \omit && \omit && \omit& \cr
\tablerule
 height2pt &\omit && \omit && \omit && \omit && \omit && \omit && \omit
&&
            \omit && \omit && \omit && \omit && \omit && \omit& \cr
  & $6_3$ && $1$  && $0$ && $7$ && $-7$
         && $0$  && $0$ && $0$ && $511$
         && $209$ &&  $-929$ && $65$ && $-449$ & \cr
 height2pt &\omit && \omit && \omit && \omit && \omit && \omit && \omit
&&
            \omit && \omit&& \omit && \omit && \omit && \omit& \cr
\tablerule  }}
\centerline{Table I} }

The second row in Table I corresponds to the unknot, which has been
labeled by $0_1$.  This, as well as the second row, which corresponds to
the trefoil, can be thought as the choice of normalization. Once those
are selected, the knot invariants $\beta_{ij}$ for all other knots are
fixed and are independent of the normalization of the Lie algebra
generators. For knots which are chiral (not amphicheiral) we have
selected in Table I the one which has $\beta_{3,1}>0$. Notice that, as
discussed before, for the two amphicheiral knots in the Table I, $4_1$
and $6_3$, one has $\beta_{ij}=0$ for $i$ odd. The values
$\beta_{2,1}(3_1)$, $\beta_{2,1}(4_1)$ and $\beta_{2,1}(5_1)$ can be
checked with the explicit computation of \alfadosunoint\ in [\gmmdos].
After taking into account the normalizations used in this work one finds
full agreement. It is important to remark that the quantities
$\beta_{ij}(K)$ are intrinsic to the knot,
\ie, as it follows from their construction they are framing independent.

In Table I only the simple  knot invariants have been listed. Vassiliev
invariants form an algebra [\birman] whose structure in the context of
this work is represented by relations \uvados. The compound knot
invariants are given by those relations from the values in Table I. The
invariants constructed in this paper allow to build actuality tables
using
\axiomone. There is one actuality table, or one Vassiliev invariant,
associated to each $\beta_{ij}$, being this simple or compound. At a
given type $i$, the space of Vassiliev invariants has dimension $m_i$,
which at least up to $i=7$ equals the quantity $d_i$ [\drortopo]. On the
other hand, we have a way to generate $d_i$ sets of Vassiliev invariants
or actuality tables. One could ask if the knot invariants constructed
are independent or not, \ie, if at each order they constitute a basis of
Vassiliev invariants. Up to order 5 this holds. At order 6 we do not
have enough information to come to  a conclusion. The invariants should
be computed for a higher number of knots. We would like, however, to
conjecture that at each order $i$, the knot invariants proposed in this
paper constitute a basis of Vassiliev invariants of type $i$.

The values of the Vassiliev invariants presented in Table I can be
contrasted with the ones given in [\vassitres]. There, knots up to 7
crossings  are considered, and Vassiliev invariants are presented up to
order 4. It turns out that $\beta_{2,1}$ and
$\beta_{3,1}$ are the same as in Table I. On the other hand, the three
invariants of order 4 seem rather different. To compare two presentations
of Vassiliev invariants of a given order one must take into account two
facts. First, they might be presented in different basis; second to a
Vassiliev invariant of order $i$ one can substract a lower order Vassiliev
invariant and still have a Vassiliev invariant of order $i$. In order to
compare our results in Table I to Vassiliev's  in [\vassitres] we must
then ask the following question:  there exist a non-singular  $3\times 3$
matrix  $A$ (which represent a change of basis) and a 3-vector $b$ such
that
$$
\pmatrix{a_{11} & a_{12} & a_{13} \cr
         a_{21} & a_{22} & a_{23} \cr
         a_{31} & a_{32} & a_{33} \cr}
\pmatrix{\beta_{4,1} \cr
         \beta_{4,2} \cr
         \beta_{4,3} \cr}
-\beta_{2,1}\pmatrix{b_1 \cr
                     b_2 \cr
                     b_3 \cr}
\eqn\comparing
$$
are the Vassiliev invariants of order 4 given in [\vassitres]?
The answer is positive. From Table I and the results in [\vassitres] one
constructs
$3\times 7=21$ linear equations whose unknowns are the 9 matrix elements
of $A$ and the 3 entries of $b$.  Remarkably, the linear system has a
solution:
$$
A={1\over 24}\pmatrix{-{12} & {1} & -{2} \cr
                0      &        0    &   {1} \cr
             {12} &   0      & -{2} \cr},
\qquad\qquad
b={1\over 24}\pmatrix{{9} \cr {5} \cr {2}\cr}.
\eqn\comres
$$
This fact is a highly non-trivial check which supports our construction.

\endpage

\chapter{Final remarks}

In this paper we have presented new numerical knot invariants which
generalize the one found in [\gmmdos]. These invariants are defined from
the universal form of the perturbative series expansion of a Wilson line
in Chern-Simons gauge theory. One of the essential ingredients to be
able to define this universal form of the series and therefore of the
invariants is the identification of all diagrams which contribute to
framing. This was done in [\pert]. In this way one works in the standard
framing and defines invariants intrinsic to knots.

Besides being framing independent, the new knot invariants have very
interesting features. First, they possess integral expressions and
therefore have a geometrical origin. Second they are of finite type, or
Vassiliev invariants, and there seems to exist a normalization in which
they are integer-valued. Third, they are computable using information on
polynomial invariants.

In this paper we have shown explicitly how the new numerical knot
invariants are computed up to order 6.  To carry out  further  the
calculation of these numerical invariants one must first develop  the
analysis of independent group structures beyond order six. In sect. 3 we
have presented an algorithm which computes these structures once the
independent Casimirs are known. We need therefore to know the form of
the independent Casimirs beyond order six.  The characterization of the
general form of these Casimirs is one of the open problems left for
future work. Once the Casimirs are known there is still the problem of
their calculation.  For the representations and Lie algebras considered
in this paper the calculation procedure described in appendix A can be
applied to any order. For other representations and groups one could use
similar techniques.

This work opens a variety of investigations. Certainly, the most
important question that one would like to answer is if the infinite
sequence of numerical invariants associated to each knot is a complete
invariant, \ie, if there are no inequivalent unoriented (invertible)
knots such that their corresponding sequences are the same. In case this
were true, one would like to know what is the minimum order (or type)
such that invariants up to that order are able to distinguish all
inequivalent unoriented knots. According to  Table I in the previous
section, for prime knots up to six crossings such a minimum order is 3.

Another important aspect which is worth to investigate is the relation
of these knot invariants with Kontsevich's approach to Vassiliev
invariants [\konse].  Both approaches should be basically the same in
the situation in which Kontsevich's is defined. Though both are integral
representations, our formulation is intrinsically three-dimensional.
This opens the possibility to define integral representations for
arbitrary three-manifolds. In this sense, one would like to search what
is the situation for other manifolds and not just for $S^3$, which is
the case considered in this paper. The new knot invariants might provide
a powerful tool to define Vassiliev invariants in the general case. One
would like also to ask if they are integer-valued  on an arbitrary
compact boundaryless manifold.

The new numerical knot invariants presented in this paper are
introduced  through the general power series expansion of  (semisimple)
group polynomial invariants. For particular groups these polynomial
invariants satisfy skein relations. It would be very interesting to
study if it is possible to write down a universal skein relation, valid
for any semisimple group. If this were possible one would have a very
useful tool to compute recursively the numerical knot invariants. Of
course, to carry this out one needs first to define numerical invariants
for links. This, on the other hand, could provide new insights on
Vassiliev invariants for links, whose theory is  much less developed
than for the case of knots. We intend to pursue these investigations in
future work.

Finally, one would like to know if the knot invariants corresponding to
each order $i$ are independent and therefore constitute a basis of
Vassiliev invariants of type $i$. If this were true (as we conjecture)
the space of quantum group invariants (at least with semisimple groups)
is the same as the space of Vassiliev invariants. This would imply in
particular that Vassiliev invariants would not be able to distinguish
non-invertible knots, since Wilson lines are invariant under
invertibility. A necessary condition for this to hold is that the
quantity $d_i$ must coincide with
$m_i$, the dimension of the space of Vassiliev invariants of type $i$.
This has been proven to be true up to order 7 [\drortopo,\birman].

\vskip1cm

\ack We would like to thank D. Bar-Natan, J. Mateos and A. V. Ramallo
for very helpful discussions.  This work was supported in part by DGICYT
under grant PB90-0772 and by CICYT under grants AEN93-0729 and
AEN94-0928.

\endpage

\Appendix{A}

In this appendix we present a summary of our group-theoretical
conventions. We choose the generators of the Lie algebra $A$ to be
antihermitian such that
$$ [T^a,T^b] = - f^{ab}_{\,\,\,\,\, c}T^c,
\eqn\auno
$$ where $f^{ab}_c$ are the structure constants.  These satisfy the
Jacobi identity,
$$
f^{ab}_{\,\,\,\,\,e}f^{ec}_{\,\,\,\,\,d}+f^{cb}_{\,\,\,\,\,e}
f^{ae}_{\,\,\,\,\,d}
+f^{ac}_{\,\,\,\,\,e}f^{be}_{\,\,\,\,\,d}=0
\eqn\jacobi
$$ The generators are normalized in such a way that for the fundamental
representation,
$$
\tr(T_aT_b)=-{1\over 2} \delta_{ab},
\eqn\normagene
$$ where $\delta_{ab}$ is the Kronecker delta. This can always be done
for  compact semisimple Lie algebras which is the case considered in
this paper.

The generators $T^a$ in the adjoint representation coincide with the
structure constants,
$$
\big(T^a\big)_c{}^b=f^{ab}_{\,\,\,\,\,c}\qquad {\hbox{\rm (adjoint
representation).}}
\eqn\adjunta
$$ The quadratic Casimir in the adjoint representation, $C_A$, is
defined as
$$ f^{ad}_{\,\,\,\,\,c} f^{bc}_{\,\,\,\,\,d}=C_A \delta^{ab}.
\eqn\Killingone
$$ The value of $C_A$ for the groups $SU(N)$ and $SO(N)$ is $-N$ and
$-{1\over 2}(N-2)$ respectively. The Killing metric is chosen to be the
identity matrix and therefore one can lower and raise group indices
freely.  For the case under consideration
$f_{abc}$ is totally antisymmetric.

The convention chosen in \auno\ seems unusual but it is the
most convenient
when  the Wilson line is defined as in \holonomy. If we  had chosen
$if^{abc}$ instead of $-f^{abc}$, the exponential of the Wilson line
would have had $ig$ instead of $g$. Our convention also introduces a
$-1$ in the vertex (see Fig. \figfeynman).

In the rest of this appendix we will describe how Casimirs are evaluated.
For the fundamental representation they are computed by means of an
algorithm presented in \REF\Cvit{P. Cvitanovi\'c, {\it Group Theory},
Nordita Classics Illustrated, 1984}
\REF\cita{P. Cvitanovi\'c\journal\pr&D14(76)1536} [\Cvit,\cita]. The
evaluation procedure is simple. We think of the Casimir as a Feynman
diagram, and introduce Feynman diagrammatic notation to replace the
algebraic expressions. The first step is to get rid of the structure
constants by means of the commutation relations
\ie, every factor like $f_{abc}T_c$ is substituted by $[T_b, T_a]$. Once
this has been done the expression to be evaluated is a linear
combination of traces of products of generators, with all their indices
contracted. For example in the calculation of $C_5$ we get, among
others, the following trace:
$$
\tr\big(T_aT_bT_cT_dT_bT_eT_dT_aT_eT_c\big).
\eqn\example
$$ In general, generators with the same index are not multiplied. If we
write the matrix elements explicitly we can put together the pairs of
generators with the same index, and therefore the quantities we are led
to evaluate are of the form,
$$
\Big(T^a\Big)_i^{\,\,\,j}\Big(T^a\Big)_k^{\,\,\,l}.
\eqn\projector
$$ These group-theoretical objects are called projection operators. They
are explicitly known for every classical Lie group except
$E_8$ [\Cvit,\cita]. In the case of $SU(N)$ the projection operators are:
$$
\Big(T^a\Big)_i^{\,\,\,j}\Big(T^a\Big)_k^{\,\,\,l}=-{1\over
2}\Big(\delta_i^{\,\,\,l}\delta_k^{\,\,\,j}-{1\over
N}\delta_i^{\,\,\,j}\delta_k^{\,\,\,l}
\Big),
\eqn\rulesun
$$ and in the case of $SO(N)$,
$$
\Big(T^a\Big)_i^{\,\,\,j}\Big(T^a\Big)_k^{\,\,\,l}=-{1\over
4}\Big(\delta_k^{\,\,\,j}\delta_i^{\,\,\,l}-\delta^{jl}\delta_{ik}\Big).
\eqn\ruleson
$$
Similar identities can be read from [\Cvit] for other
groups. This solves the problem of calculating the Casimirs in the
fundamental representation of these groups.

Higher representations can be introduced as properly symmetrized
products of fundamental representations. These products span the
representation ring of any compact Lie group. Extensions of \rulesun\
and \ruleson\ can be found, which would enable us to evaluate the analog
of \projector\ in these more involved cases. Nevertheless in the case of
$SU(2)$ in representation $j$ nothing of this is needed since its rank
is 1 and therefore all Casimirs are linear combinations of powers of the
quadratic Casimir $j(j+1)$. The key identity is the product of two
structure constants (which will be denoted by $\varepsilon_{ijm}$) with
only one index contracted,
$$
\varepsilon_{ijm}\varepsilon_{mkl}=\delta_{ik}\delta_{jl}-\delta_{il}
\delta_{jk}.
\eqn\epsilons
$$ Clearly this well-known identity simplifies the calculation of
Casimirs in the case of $SU(2)$. Any Casimir gets reduced to the
quadratic Casimir in the corresponding representation.

Using these rules we have computed all the Casimirs up to order six
which have been used in this paper. These correspond to the fundamental
representations of
$SU(N)$ and  $SO(N)$,  and to an arbitrary spin $j$ representation of
$SU(2)$. Their values are contained in the following list:
$$
\eqalign{ SU(N)_f: \qquad C_2=& -{1\over 2N}(N^2-1)\cr
                C_3=& -{1\over 4}(N^2-1)\cr
                C_4=& {1\over 16}(N^2-1)(N^2+2)\cr
                C_5=& {1\over 32}N(N^2-1)(N^2+1)\cr
                C_6^1=& {1\over 64}(N^2-1)(N^4+N^2+2)\cr
                C_6^2=& {1\over 64}(N^2-1)(3N^2-2)\cr}
                \eqn\casivalsun
$$
$$\eqalign{  SO(N)_f: \qquad C_2=& -{1\over 4}(N-1)\cr
                C_3=& -{1\over 16}(N-1)(N-2)\cr
                C_4=& {1\over 256}(N-1)(N-2)(N^2-5N+10)\cr
                C_5=& {1\over 1024}(N-1)(N-2)(N^3-7N^2+17N-10)\cr
                C_6^1=& {1\over 4096}(N-1)(N-2)(N^2-7N+14)(N^2-2N+3)\cr
                C_6^2=& {1\over 4096}(N-1)(N-2)(N-3)(7N-18)\cr}
                \eqn\casivalson
$$
$$\eqalign{  SU(2)_j: \qquad C_2=& -j(j+1)\cr
                C_3=& -j(j+1)\cr
                C_4=& 2j^2(j+1)^2\cr
                C_5=& 3j^2(j+1)^2-j(j+1) \cr
                C_6^1=& 2j^3(j+1)^3+3j^2(j+1)^2-2j(j+1) \cr
                C_6^2=& -2j^3(j+1)^3+5j^2(j+1)^2-2j(j+1)\cr}
\eqn\casivalsuj
$$

\endpage

\Appendix{B} In this appendix we present some elementary facts about
semisimple Lie algebras relevant to the analysis of group factors in the
perturbative expansion.  Consider the part of a diagram depicted on the
left of \FIG\figunonum\ Fig. \figunonum . It is possible to reduce its
group factor by means of the totally antisymmetry of the structure constants
$f_{abc}$ at the cost of introducing $C_A$:
$$ f_{abc}T_bT_c =\half f_{abc}\big[ T_b, T_c\big]=\half
f_{abc}f_{cbd}T_d =\half C_A T_a.
\eqn\identity
$$ Actually, the Casimir $C_A$ can be written in terms of $C_2$ and
$C_3$, which is the expression that we will use wherever a ``fishtail''
appears:
$$ C_3= -{1\over d(R)}f_{abc}\tr (T_aT_bT_c)=-{1\over 2d(R)} C_A
\tr(T_aT_a)=-\half C_A C_2
\quad\Rightarrow\quad C_A=-{2C_3\over C_2}.
\eqn\identitytwo
$$

There is a point which may be worth commenting. For an arbitrary
semisimple Lie algebra
$A=\oplus_{k=1}^n A_k$ in an arbitrary representation, the Wilson line
can be imagined as consisting of $n$ Wilson lines each one corresponding
to one of the simple Lie algebras $A_k$ in its respective
representation. Therefore, a connected (sub)diagram  can be regarded as
a sum of similar subdiagrams where in each term the legs are attached to
a component of the Wilson line and the sum runs over all these
components. As a consequence one cannot identify the group factors of
diagrams like those in
\FIG\differents\ Fig. \differents\ due to crossed terms in the product
of the group factors of the subdiagrams. If the Lie algebra were simple
their group factor would be the same due to the uniqueness of the Wilson
line and to the reduction of the fishtails.

We call a subdiagram ``separated'' if its endpoints do not enclose any
endpoints belonging to another subdiagram. For example, the diagrams
displayed in Fig. \differents\ include only separated subdiagrams. In
these cases the group factor is a product of the group factors of the
subdiagrams.

Once a diagram has been drawn, its group factor is found in the following
fashion. First one has to transform it into a sum of diagrams with only
separated connected subdiagrams by means of the commutation relations.
Take one of these new diagrams and consider one of its connected
subdiagrams. Then reduce all the ``fishtails'' of the subdiagram chosen
by means of
\identity\ until one finishes with a Casimir-like subdiagram. The group
factor of these Casimir-like subdiagrams has to be calculated separately
and written in terms of its corresponding Casimirs by repeated use of
the commutation relations. Then sum over all algebras $A_k$. This would
give the contribution corresponding to the chosen subdiagram. Repeat
this procedure for all subdiagrams and multiply the contributions of
each subdiagram. The result is the group factor of the diagram we begun
with. Following these steps for all diagrams present in the sum, we are
done.

For example, the group factors of Fig. \differents\ are readily found to
be
$$
\eqalign{
r(D_1)=&\Big(\sum_{k=1}^n\big(C_3^{(k)}\big)^2
\big(C_2^{(k)}\big)^{-1}\Big)^2
=r_{3,1}^2,\cr r(D_2)=&\Big(\sum_{k=1}^n
C_3^{(k)}\Big)\sum_{k=1}^n\big(C_3^{(k)}\big)^3\big(C_2^{(k)}\big)^{-2}
=r_{2,1}r_{4,2}.\cr}
\eqn\ultima
$$ Notice that for a simple algebra these two factors would be the same.
However, in the semisimple case they are different.

\endpage

\Appendix{C}

In this appendix we list the polynomial knot invariants which have been
used in this paper to compute the numerical knot invariants up to order
six for prime knots up to six crossings. Knot polynomials for the case
of $SU(2)$ in a representation of spin $j$ are collected from [\kguno].
They are:
$$
\eqalign{ 0_1:\qquad V_j=& 1, \cr  3_1:\qquad V_j=&{1\over [2j+1]}
\sum_{l=0}^{2j}[2l+1](-1)^{2j-l}q^{-3(2j(2j+2)-l(l+1))/2}, \cr
4_1:\qquad V_j=&{1\over [2j+1]}
\sum_{l,k=0}^{2j}\sqrt{[2l+1][2k+1]}a_{kl}q^{l(l+1)-k(k+1)},\cr
5_1:\qquad V_j=&{1\over [2j+1]}
\sum_{l=0}^{2j}[2l+1](-1)^{2j-l}q^{-5(2j(2j+2)-l(l+1))/2}, \cr
5_2:\qquad V_j=& {1\over [2j+1]}
\sum_{l,k=0}^{2j}\sqrt{[2l+1][2k+1]}a_{kl}(-1)^k
q^{2j(2j+2)-l(l+1)+3k(k+1)/2},\cr 6_1:\qquad V_j=&{1\over [2j+1]}
\sum_{l,k=0}^{2j}\sqrt{[2l+1][2k+1]}a_{lk}q^{l(l+1)-2k(k+1)},\cr
6_2:\qquad V_j=&{1\over [2j+1]}
\sum_{i,l,k=0}^{2j}\sqrt{[2i+1][2k+1]}a_{li}a_{lk}(-1)^ {2j-l-k} \cr
&q^{-3(2j(2j+2)-k(k+1))/2-l(l+1)/2+i(i+1)},\cr 6_3:\qquad V_j=&{1\over
[2j+1]}
\sum_{r,s,l,k=0}^{2j}\sqrt{[2s+1][2k+1]}a_{kl}a_{lr}a_{rs}(-1)^ {l+r}\cr
& \qquad \qquad \qquad q^{-k(k+1)+s(s+1)+l(l+1)/2-r(r+1)/2},\cr}
\eqn\freson
$$ where,
$$ [m]={{q^{m/2}-q^{-m/2}}\over {q^{1/2}-q^{-1/2}}},
\eqn\mango
$$ is a $q$-number,
$$ [m]!=[m][m-1][m-2]\ldots [2][1],
\eqn\manga
$$ corresponds to the  $q$-factorial of a $q$-number, and,
$$ a_{kl}=(-1)^{l+k-2j}\sqrt{[2k+1][2l+1]}\left(\matrix{j & j & k \cr j &
j & l\cr}\right),
\eqn\limonero
$$ is the duality matrix. This matrix is given in terms of the quantum
Racah coefficients,
$$
\eqalign{
\left(\matrix{j_1 & j_2 & j_{12} \cr j_3 & j_4 &
j_{23}\cr}\right)=&\Delta(j_1,j_2,j_{12})
\Delta(j_3,j_4,j_{12})\Delta(j_1,j_4,j_{23})
\Delta(j_3,j_2,j_{23})\cr&\sum_{m\geq0}(-1)^m[m+1]!
\big{\{}[m-j_1-j_2-j_{12}]!
[m-j_3-j_4-j_{12}]!\cr&[m-j_1-j_4-j_{23}]![m-j_3-j_2-j_{23}]!
[j_1+j_2+j_3+j_4-m]!\cr&
[j_1+j_3+j_{12}+j_{23}-m]![j_2+j_4+j_{12}+j_{23}-m]!\big{\}}^{-1},\cr}
\eqn\platano
$$ where the sum over $m$ runs over all non-negative integers such that
none of the
$q$-factorials get a negative argument, and the symbol
$\Delta(a,b,c)$ stands for
$$
\Delta(a,b,c)=\sqrt{{[-a+b+c]![a-b+c]![a+b-c]!}\over{[a+b+c+1]!}}.
\eqn\libro
$$

For $SU(N)$ one has the following list of HOMFLY polynomial invariants
\REF\wuwu{F.Y. Wu\journal\rmp&64(92)1099} [\homflyp,\jonesAM,\wuwu]:
$$
\eqalign{ 0_1: \qquad H  & = 1, \cr 3_1: \qquad H  & = {\lambda}\,\left(
1 + {q^2} - {\lambda}\,{q^2} \right), \cr 4_1: \qquad H  & = 1 - {1\over
q} + {1\over {{\lambda}\,q}} - q + {\lambda}\,q, \cr 5_1: \qquad H  & =
{{\lambda}^2}\,\left( 1 + {q^2} - {\lambda}\,{q^2} + {q^4} -
{\lambda}\,{q^4} \right), \cr 5_2: \qquad H  & = {\lambda}\,\left( 1 - q
+ {\lambda}\,q + {q^2} - {\lambda}\,{q^2} + {\lambda}\,{q^3} -
{{\lambda}^2}\,{q^3} \right), \cr 6_1: \qquad H  & =
{{{\lambda}^{-2}}\,{q^{-2}}}({1 - {\lambda} + {\lambda}\,q -
{{\lambda}^2}\,q - {\lambda}\,{q^2} + 2\,{{\lambda}^2}\,{q^2} -
{{\lambda}^2}\,{q^3} +
     {{\lambda}^3}\,{q^3}}), \cr 6_2: \qquad H  & =
{{{\lambda}^{-2}}\,{q^{-3}}} ({1 - {\lambda} - q + {\lambda}\,q + {q^2}
- 2\,{\lambda}\,{q^2} + {{\lambda}^2}\,{q^2} + {\lambda}\,{q^3} -
{\lambda}\,{q^4} +
     {{\lambda}^2}\,{q^4}}) , \cr 6_3: \qquad H  & = 3 - {1\over
{\lambda}} - {\lambda} + {q^{-2}} - {1\over {{\lambda}\,{q^2}}} -
{1\over q} + {1\over {{\lambda}\,q}} -
  q + {\lambda}\,q + {q^2} - {\lambda}\,{q^2}, \cr}
$$ where $\lambda=q^{N-1}$.

Finally, for the Kauffman polynomial invariants [\kauf,\wuwu]:
$$
\eqalign{ 3_1: \quad R & = 2 {{\alpha}^2} - {{\alpha}^4} + \left(
-{{\alpha}^3} + {{\alpha}^5} \right)  z +
  \left( {{\alpha}^2} - {{\alpha}^4} \right)  {z^2} \cr 4_1: \quad R & =
-1 + {{\alpha}^{-2}} + {{\alpha}^2} + \left( {1\over {\alpha}} -
{\alpha} \right)  z +
  \left( -2 + {{\alpha}^{-2}} + {{\alpha}^2} \right)  {z^2} +
  \left( {1\over {\alpha}} - {\alpha} \right)  {z^3} \cr 5_1: \quad R & =
3 {{\alpha}^4} - 2 {{\alpha}^6} + \left( -2 {{\alpha}^5} + {{\alpha}^7}
+ {{\alpha}^9} \right)
   z + \left( 4 {{\alpha}^4} - 3 {{\alpha}^6} - {{\alpha}^8} \right)
{z^2} +
  \left( -{{\alpha}^5} + {{\alpha}^7} \right)  {z^3} \cr & \quad  +
  \left( {{\alpha}^4} - {{\alpha}^6} \right)  {z^4} \cr 5_2: \quad R & =
{{\alpha}^2} + {{\alpha}^4} - {{\alpha}^6} + \left( -2 {{\alpha}^5} + 2
{{\alpha}^7} \right)  z +
  \left( {{\alpha}^2} + {{\alpha}^4} - 2 {{\alpha}^6} \right)  {z^2} +
  \left( {{\alpha}^3} - 2 {{\alpha}^5} + {{\alpha}^7} \right)  {z^3} \cr
& \quad  +
  \left( {{\alpha}^4} - {{\alpha}^6} \right)  {z^4} \cr 6_1: \quad R & =
{{\alpha}^{-4}} - {{\alpha}^{-2}} + {{\alpha}^2} +
  \left( {2\over {{{\alpha}^3}}} - {2\over {\alpha}} \right)  z +
  \left( {3\over {{{\alpha}^4}}} - {4\over {{{\alpha}^2}}} +
{{\alpha}^2} \right)
   {z^2} + \left( {3\over {{{\alpha}^3}}} - {2\over {\alpha}} - {\alpha}
\right)
   {z^3}\cr & \quad  + \left( 1 + {{\alpha}^{-4}} - {2\over
{{{\alpha}^2}}} \right)
   {z^4}  + \left( {{\alpha}^{-3}} - {1\over {\alpha}} \right)  {z^5} \cr
6_2: \quad R & = 2 + {{\alpha}^{-4}} - {2\over {{{\alpha}^2}}} +
  \left( {{\alpha}^{-5}} - {{\alpha}^{-3}} \right)  z +
  \left( 3 + {{\alpha}^{-6}} + {2\over {{{\alpha}^4}}} - {6\over
{{{\alpha}^2}}}
\right)  {z^2}  + \left( {2\over {{{\alpha}^5}}} - {2\over {\alpha}}
\right)  {z^3}
\cr & \quad  + \left( 1 + {2\over {{{\alpha}^4}}} -
     {3\over {{{\alpha}^2}}} \right)  {z^4}   +
  \left( {{\alpha}^{-3}} - {1\over {\alpha}} \right)  {z^5} \cr 6_3:
\quad R & = 3 - {{\alpha}^{-2}} - {{\alpha}^2} + \left( -{{\alpha}^{-3}}
+ {2\over {\alpha}} - 2 {\alpha} +
     {{\alpha}^3} \right)  z + \left( 6 - {3\over {{{\alpha}^2}}} -
     3 {{\alpha}^2} \right)  {z^2} \cr & \quad  +
  \left( -{{\alpha}^{-3}} + {1\over {\alpha}} - {\alpha} + {{\alpha}^3}
\right)  {z^3} +
  \left( 4 - {2\over {{{\alpha}^2}}} - 2 {{\alpha}^2} \right)  {z^4} +
  \left( -{1\over {\alpha}} + {\alpha} \right)  {z^5}\cr}
$$ where $z=q^{1\over 4}-q^{-{1\over 4}}$ and $\alpha = q^{N-1\over 4}$.
These invariant polynomials, as well as the rest of the invariants of
this appendix, are given in the standard framing.

\endpage

\refout
\endpage
\end

\ifigc\figfeynman{Feynman rules.}{figfeynman.ps}{8}{4}
\ifigc\figresults{One-loop diagrams.}{figresults.ps}{4}{4}
\ifigc\figvarios{Feynman diagrams at order three.}{figvarios.ps}{7}{4.55}
\ifigc\figtresnum{Diagrams associated to  Casimirs
$C_2$, $C_3$, $C_4$, $C_5$, $C_6^1$ and $C_6^2$.}{casimiros.ps}{6}{2.5}
\ifigc\generaldiag{A generic diagram.}{generaldiag.ps}{4}{6.3}
\ifigc\freeprop{A  generic diagram including a collapsible
propagator.}{freeprop.ps}{4}{6}
\ifigc\generalvert{A generic diagram including a
vertex.}{generalvert.ps}{4}{6.3}
\ifigc\generalfish{Diagrams coming from the  previous
figure.}{generalfish.ps}{4}{4.2}
\ifigc\casimirlike{Examples of connected and disconnected Casimir-like
diagrams.}{casimirlike.ps}{4}{4}
\ifigc\casson{Diagrams contributing to $\alpha_{2,1}$.}{casson.ps}{4}{4.5}
\ifig\laserres{Representative diagrams for the group factors $r_{ij}$ for
$i=2,\cdots,6$.}{erressave.ps}{14}{3.5}
\ifigc\crossings{Extension of $V(K)$ to singular
knots}{crossings.ps}{2}{4}
\ifigc\figunonum{Reduction of a ``fishtail''.}{figunonum.ps}{3}{4.3}
\ifigc\differents{Diagrams with different group
factors.}{differents.ps}{4}{4}

\end